# Compact Formulation of the First Evolution Equation for Optimal Control Computation

Sheng ZHANG, Fei LIAO, and Wei-Qi QIAN

(2018.03)

*Abstract:* **The first evolution equation is derived under the Variation Evolving Method (VEM) that seeks optimal solutions with the variation evolution principle. To improve the performance, its compact form is developed. By replacing the states and costates variation evolution with that of the controls, the dimension-reduced Evolution Partial Differential Equation (EPDE) only solves the control variables along the variation time to get the optimal solution, and its definite conditions may be arbitrary. With this equation, the scale of the resulting Initial-value Problem (IVP), transformed via the semi-discrete method, is significantly reduced. Illustrative examples are solved and it is shown that the compact form evolution equation outperforms the primary form in the precision, and the efficiency may be higher for the dense discretization. Moreover, in discussing the connections to the classic iteration methods, it is uncovered that the computation scheme of the gradient method is the discrete implementation of the third evolution equation, and the compact form of the first evolution equation is a continuous realization of the Newton type iteration mechanism.**

## I. INTRODUCTION

Optimal control theory aims to determine the inputs to a dynamic system that optimize a specified performance index while satisfying constraints on the motion of the system. It is closely related to the engineering and has been widely studied [1]. Because of the complexity, Optimal Control Problems (OCPs) are usually solved with numerical methods. Various numerical methods are developed and generally they are divided into two classes, namely, the direct methods and the indirect methods [2]. The direct methods discretize the control or/and state variables to obtain the Nonlinear Programming (NLP) problem, for example, the widely-used direct shooting method [3] and the classic collocation method [4]. These methods are easy to apply, whereas the results obtained are usually suboptimal [5], and the optimal may be infinitely approached. The indirect methods transform the OCP to a Boundary-value Problem (BVP) through the optimality conditions. Typical methods of this type include the well-known indirect shooting method [2] and the novel symplectic method [6]. Although be more precise, the indirect methods often suffer from the significant numerical difficulty due to the ill-conditioning of the Hamiltonian dynamics, that is, the stability of the costates dynamics is adverse to that of the states dynamics [7]. The recent development, representatively the Pseudo-spectral (PS) method [8], blends the two types of methods, as it unifies the direct computation of NLP and the indirect realization of BVP in a dualization view [9]. Such methods inherit the advantages of both types and blur their difference.

Theories in the dynamics and control field often enlighten strategies for the optimal solutions, for example, the non-linear

The authors are with the Computational Aerodynamics Institution, China Aerodynamics Research and Development Center, Mianyang, 621000, China. (e-mail: zszhangshengzs1@outlook.com).

variable transformation to reduce the variables [10] and the dynamic method to solve the unconstrained parameter optimization problems [11]. Recently, a Variation Evolving Method (VEM), inspired by the states evolution within the stable continuous-time dynamic system, is proposed for the optimal control computation [12]-[20]. The VEM is built upon the infinite-dimensional dynamics Lyapunov principle, and it synthesizes the direct and indirect methods from a new standpoint. The Evolution Partial Differential Equation (EPDE), which describes the evolution of variables towards the optimal solution, is derived from the viewpoint of variation motion to reduce the performance index, and the optimality conditions will be gradually met with theoretical guarantee. In Refs. [12] and [13], besides the states and the controls, the costates are also employed in developing the EPDE (also named the first evolution equation), and this increases the complexity of the computation. In Refs. [14]-[19], the VEM that uses only the original variables is systematically developed. The costate-free optimality conditions are established and the second evolution equation is derived. In Ref. [20], the third evolution equation, the compact form of the second evolution equation in essence, is proposed. The EPDE therein only concerns the control variables and hence relieves the computation burden. To enhance the performance with the similar manner, in this paper the compact formulation of the first evolution equation will be studied.

Throughout the paper, our work is built upon the assumption that the solution for the optimization problem exists. We do not describe the existence conditions for the purpose of brevity. Relevant researches such as the Filippov-Cesari theorem are documented in Ref. [21]. In the following, first the principle of the VEM is reviewed. Then the first evolution equation, which addresses typical OCPs with terminal constraints, is presented. The compact formulation stems from a more concise unconstrained functional, and its form is quite different from the primary form. Later, illustrative examples are solved to test the developed equation. We also discuss the connections of the evolution equations to the classis iteration numerical methods at the end.

## II. Preliminaries

### A. Infinite-dimensional Lyapunov theory

The VEM is a newly developed method for the optimal solutions. It is enlightened from the inverse consideration of the Lyapunov dynamics stability theory in the control field [22]. As the start point of this method, the finite-dimensional Lyapunov principle is generalized for the infinite-dimensional continuous-time dynamics.

**Definition 1**: For an infinite-dimensional dynamic system described by

$$\frac{\partial \boldsymbol{y}(x,t)}{\partial t} = \boldsymbol{f}(\boldsymbol{y},x) \tag{1}$$

where $t$ is the time. $x \in \mathbb{R}$ is the independent variable. $\boldsymbol{y}(x) \in \mathbb{D}(x) \in \mathbb{R}^n(x)$ is the function vector of $x$. $\boldsymbol{f}: \mathbb{D}(x) \times \mathbb{R} \to \mathbb{R}^n(x)$ is a vector function. $\mathbb{D}(x)$ is a certain function set. If $\hat{\boldsymbol{y}}(x) \in \mathbb{D}(x)$ satisfies $\boldsymbol{f}(\hat{\boldsymbol{y}}(x),x) = \boldsymbol{0}$, then $\hat{\boldsymbol{y}}(x)$ is called an equilibrium solution.

**Definition 2**: The equilibrium solution $\hat{\boldsymbol{y}}(x)$ is an asymptotically stable solution in $\mathbb{D}(x)$ if for any initial conditions $\boldsymbol{y}(x,t)\big|_{t=0} = \tilde{\boldsymbol{y}}(x) \in \mathbb{D}(x)$, there is $\lim_{t \to +\infty} \left\| \boldsymbol{y}(x,t) - \hat{\boldsymbol{y}}(x) \right\|_\infty = 0$, where $\left\| \boldsymbol{y} \right\|_\infty = \sup_x \left( \sum_{i=1}^{n} y_i^2 \right)^{1/2}$ denote the vector function supremum norm.

**Lemma 1**: For the infinite-dimensional dynamic system (1), if there exists a continuously differentiable functional $V: \mathbb{D}(x) \to \mathbb{R}$ such that

i) $V(\hat{\boldsymbol{y}}(x)) = c$ and $V(\boldsymbol{y}(x)) > c$ in $\mathbb{D}(x)/\{\hat{\boldsymbol{y}}(x)\}$.

ii) $\dot{V}(\boldsymbol{y}(x)) \le 0$ in $\mathbb{D}(x)$ and $\dot{V}(\boldsymbol{y}(x)) < 0$ in $\mathbb{D}(x)/\{\hat{\boldsymbol{y}}(x)\}$.

where $c$ is a constant. Then $\boldsymbol{y}(x) = \hat{\boldsymbol{y}}(x)$ is an asymptotically stable solution in $\mathbb{D}(x)$.

**Proof**: To prove the lemma, we need to show that $\lim_{t \to +\infty} \left\| \boldsymbol{y}(x,t) - \hat{\boldsymbol{y}}(x) \right\|_\infty = 0$. Given any scalar $a > 0$, we can find a function set $\mathbb{W}_b$, defined as $\mathbb{W}_b = \left\{ \boldsymbol{y}(x) \,|\, V(\boldsymbol{y}(x)) \le b \right\}$ with $b = \min_{\left\| \boldsymbol{y}(x) - \hat{\boldsymbol{y}}(x) \right\|_\infty = a} V(\boldsymbol{y}(x))$, such that $\mathbb{W}_b \subset \mathbb{B}_a = \left\{ \boldsymbol{y}(x) \,|\, \left\| \boldsymbol{y}(x,t) - \hat{\boldsymbol{y}}(x) \right\|_\infty < a \right\} \subset \mathbb{D}(x)$. Therefore, the statement may be proved by showing that $\lim_{t \to +\infty} V(\boldsymbol{y}(x)) = c$. Because the functional $V$ is monotonically decreasing in $\mathbb{D}(x)/\{\hat{\boldsymbol{y}}(x)\}$ and bounded below by $c$, we have

$$\lim_{t \to +\infty} V = \hat{c} \ge c \tag{2}$$

To show that $\hat{c} = c$, we use a contradiction argument. Suppose $\hat{c} > c$. By the continuity of $V$, there is $d > 0$ such that $\mathbb{B}_d \subset \mathbb{W}_{\hat{c}}$. Eq. (2) implies that $\boldsymbol{y}(x)$ lies outside $\mathbb{B}_d$ for all $t > 0$. Let

$$\gamma = \max_{\boldsymbol{y}(x) \in \mathbb{S}} \dot{V}(\boldsymbol{y}(x)) < 0 \tag{3}$$

where the set $\mathbb{S}$ is defined as $\mathbb{S} = \left\{ \boldsymbol{y}(x) \,|\, d \le \left\| \boldsymbol{y}(x,t) - \hat{\boldsymbol{y}}(x) \right\|_\infty \le \left\| \tilde{\boldsymbol{y}}(x) - \hat{\boldsymbol{y}}(x) \right\|_\infty \right\}$. The existence of $\gamma$ is guaranteed because the continuous functional $\dot{V}$ has a maximum over the compact set $\mathbb{S}$. Thus

$$V(\boldsymbol{y}(x)) = V(\tilde{\boldsymbol{y}}(x)) + \int_0^t \dot{V} \mathrm{d}s \le V(\tilde{\boldsymbol{y}}(x)) + \gamma t \tag{4}$$

Since the right-hand side of Eq. (4) will eventually become smaller than $c$, the inequality contradicts the assumption of $\hat{c} > c$. ■

*B. Principle of VEM*

The VEM analogizes the optimal solution to the asymptotically stable equilibrium point of an infinite-dimensional dynamic system, and derives such dynamics to minimize a specific performance index that acts the Lyapunov functional. To implement the idea, a virtual dimension, the variation time $\tau$, is introduced to describe the process that a variable like $\boldsymbol{u}(t)$ evolves to the optimal solution under the dynamics governed by the variation dynamic evolution equations, which may be presented in the form of the EPDE and the Evolution Differential Equation (EDE). Fig. 1 illustrates the variation evolution of the control variables in the VEM to solve the OCP. Through the variation motion, the initial guess of control will evolve to the optimal solution, and the optimality conditions will be gradually achieved.

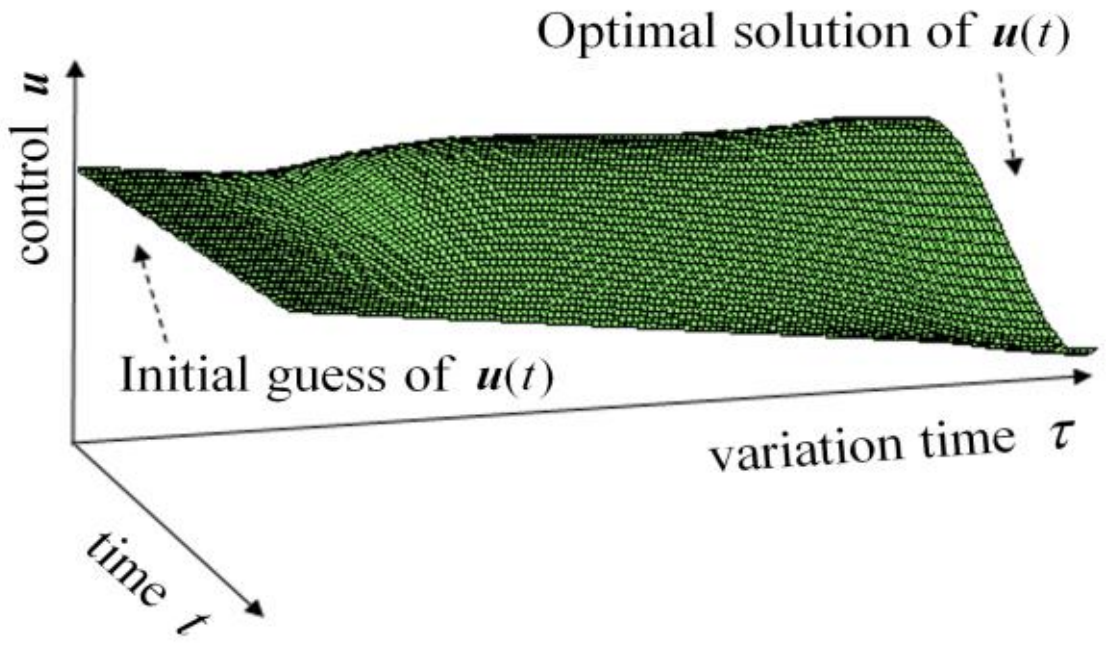

Fig. 1. The illustration of the control variable evolving along the variation time $\tau$ in the VEM.

For example, consider the calculus-of-variations problem defined as

$$J = \int_{t_0}^{t_f} F\left(\boldsymbol{y}(t), \dot{\boldsymbol{y}}(t), t\right) \mathrm{d}t \tag{5}$$

where the elements of the variable vector $\boldsymbol{y}(t) \in \mathbb{R}^n$ belong to $C^2[t_0, t_f]$. $t_0$ and $t_f$ are the fixed initial and terminal time. The EPDE and EDEs derived are [12][14]

$$\frac{\partial \boldsymbol{y}(t,\tau)}{\partial \tau} = -\boldsymbol{K}\left( F_{\boldsymbol{y}} - \frac{\partial}{\partial t}(F_{\dot{\boldsymbol{y}}}) \right) \ , t \in [t_0, t_f] \tag{6}$$

$$\frac{\mathrm{d}\boldsymbol{y}(t_0)}{\mathrm{d}\tau} = \boldsymbol{K}\, F_{\dot{\boldsymbol{y}}}\Big|_{t_0} \tag{7}$$

$$\frac{\mathrm{d}\boldsymbol{y}(t_f)}{\mathrm{d}\tau} = -\boldsymbol{K}\, F_{\dot{\boldsymbol{y}}}\Big|_{t_f} \tag{8}$$

where the column vectors $F_{\boldsymbol{y}} = \dfrac{\partial F}{\partial \boldsymbol{y}}$ and $F_{\dot{\boldsymbol{y}}} = \dfrac{\partial F}{\partial \dot{\boldsymbol{y}}}$ are the shorthand notations of partial derivatives, and $\boldsymbol{K}$ is a $n \times n$ dimensional positive-definite gain matrix. The equilibrium solution of Eqs. (6)-(8) will satisfy the optimality conditions, i.e., the Euler-Lagrange equation [23][24]

$$F_{\boldsymbol{y}} - \frac{\mathrm{d}}{\mathrm{d}t}(F_{\dot{\boldsymbol{y}}}) = \boldsymbol{0} \tag{9}$$

and the transversality conditions

$$F_{\dot{\boldsymbol{y}}}\Big|_{t_0} = \boldsymbol{0} \tag{10}$$

$$F_{\dot{\boldsymbol{y}}}\Big|_{t_f} = \boldsymbol{0} \tag{11}$$

To seek the optimal solution, we need to solve the EPDE and the EDEs with right definite conditions. Via the well-known semi-discrete method in the field of Partial Differential Equation (PDE) numerical calculation [25], those equations are transformed to the finite-dimensional Initial-value Problems (IVPs) to be solved, with the common Ordinary Differential Equation (ODE) integration methods. Note that the resulting IVPs are defined with respect to the variation time $\tau$, not the normal time $t$.

## III. The First Evolution Equations for OCP

### *A. OCP definition*

In this paper, we consider the OCPs with terminal constraint that are defined as

**Problem 1**: Consider performance index of Bolza form

$$J = \varphi(\boldsymbol{x}(t_f), t_f) + \int_{t_0}^{t_f} L\left(\boldsymbol{x}(t), \boldsymbol{u}(t), t\right) \mathrm{d}t \tag{12}$$

subject to the dynamic equation

$$\dot{\boldsymbol{x}} = \boldsymbol{f}(\boldsymbol{x}, \boldsymbol{u}, t) \tag{13}$$

where $t \in \mathbb{R}$ is the time. $\boldsymbol{x} \in \mathbb{R}^n$ is the state vector and its elements belong to $C^2[t_0, t_f]$. $\boldsymbol{u} \in \mathbb{R}^m$ is the control vector and its elements belong to $C^1[t_0, t_f]$. The function $L : \mathbb{R}^n \times \mathbb{R}^m \times \mathbb{R} \to \mathbb{R}$ and its first-order partial derivatives are continuous with respect to $\boldsymbol{x}$, $\boldsymbol{u}$ and $t$. The function $\varphi : \mathbb{R}^n \times \mathbb{R} \to \mathbb{R}$ and its first-order and second-order partial derivatives are continuous with respect to $\boldsymbol{x}$ and $t$. The vector function $\boldsymbol{f}:\mathbb{R}^n\times\mathbb{R}^m\times\mathbb{R}\rightarrow\mathbb{R}^n$ and its first-order partial derivatives are continuous and Lipschitz in $\boldsymbol{x}$, $\boldsymbol{u}$ and $t$. The initial time $t_0$ is fixed and the terminal time $t_f$ is free. The initial and terminal boundary conditions are respectively prescribed as

$$\boldsymbol{x}(t_0)=\boldsymbol{x}_0 \tag{14}$$

$$\boldsymbol{g}\left(\boldsymbol{x}(t_f),t_f\right)=\boldsymbol{0} \tag{15}$$

where $\boldsymbol{g}:\mathbb{R}^n\times\mathbb{R}\rightarrow\mathbb{R}^q$ is a $q$ dimensional vector function with continuous first-order partial derivatives. Find the optimal solution $(\hat{\boldsymbol{x}},\hat{\boldsymbol{u}})$ that minimizes $J$, i.e.

$$(\hat{\boldsymbol{x}},\hat{\boldsymbol{u}})=\arg\min(J) \tag{16}$$

*B. The primary formulation*

The first evolution equation that solves Problem 1 are derived through a constructed unconstrained functional, which employs the classic optimality conditions as

$$\begin{aligned}\bar{J}=&\left(\boldsymbol{x}(t_0)-\boldsymbol{x}_0\right)^{\mathrm{T}}\boldsymbol{W}_{\boldsymbol{x}_0}\left(\boldsymbol{x}(t_0)-\boldsymbol{x}_0\right)+\boldsymbol{g}^{\mathrm{T}}\boldsymbol{W}_{\boldsymbol{x}_f}\boldsymbol{g}+\left(\boldsymbol{\lambda}(t_f)-\varphi_{\boldsymbol{x}}(t_f)-\boldsymbol{g}_{\boldsymbol{x}}{}^{\mathrm{T}}(t_f)\boldsymbol{\pi}\right)^{\mathrm{T}}\boldsymbol{W}_{\boldsymbol{\lambda}_f}\left(\boldsymbol{\lambda}(t_f)-\varphi_{\boldsymbol{x}}(t_f)-\boldsymbol{g}_{\boldsymbol{x}}{}^{\mathrm{T}}(t_f)\boldsymbol{\pi}\right)\\&+w_H\left(H(t_f)+\varphi_t(t_f)+\boldsymbol{\pi}^{\mathrm{T}}\boldsymbol{g}_t(t_f)\right)^2+\int_{t_0}^{t_f}\left\{\left(\dot{\boldsymbol{x}}-H_{\boldsymbol{\lambda}}\right)^{\mathrm{T}}\left(\dot{\boldsymbol{x}}-H_{\boldsymbol{\lambda}}\right)+\left(\dot{\boldsymbol{\lambda}}+H_{\boldsymbol{x}}\right)^{\mathrm{T}}\left(\dot{\boldsymbol{\lambda}}+H_{\boldsymbol{x}}\right)+H_{\boldsymbol{u}}{}^{\mathrm{T}}H_{\boldsymbol{u}}\right\}\mathrm{d}t\end{aligned} \tag{17}$$

where $H=L+\boldsymbol{\lambda}^{\mathrm{T}}\boldsymbol{f}$ is the Hamiltonian, $\boldsymbol{\lambda}\in\mathbb{R}^n$ is the costates, and $\boldsymbol{\pi}\in\mathbb{R}^q$ is the Lagrange multipliers that adjoin the terminal constraint (15) in the classic adjoining method [26]. $\boldsymbol{W}_{\boldsymbol{x}_0}$, $\boldsymbol{W}_{\boldsymbol{x}_f}$ and $\boldsymbol{W}_{\boldsymbol{\lambda}_f}$ are right dimensional positive-definite weight matrixes. $w_H$ is a positive weight constant. "T" is the transpose operator. Obviously, the minimum solution of the unconstrained functional (17) satisfies the feasibility conditions (13)-(15) and the classic optimality conditions, i.e.

$$\dot{\boldsymbol{\lambda}}+H_{\boldsymbol{x}}=\dot{\boldsymbol{\lambda}}+L_{\boldsymbol{x}}+\boldsymbol{f}_{\boldsymbol{x}}{}^{\mathrm{T}}\boldsymbol{\lambda}=\boldsymbol{0} \tag{18}$$

$$H_{\boldsymbol{u}}=L_{\boldsymbol{u}}+\boldsymbol{f}_{\boldsymbol{u}}{}^{\mathrm{T}}\boldsymbol{\lambda}=\boldsymbol{0} \tag{19}$$

and the transversality conditions

$$\boldsymbol{\lambda}(t_f)-\varphi_{\boldsymbol{x}}(t_f)-\boldsymbol{g}_{\boldsymbol{x}}{}^{\mathrm{T}}(t_f)\boldsymbol{\pi}=\boldsymbol{0} \tag{20}$$

$$H(t_f)+\varphi_t(t_f)+\boldsymbol{\pi}^{\mathrm{T}}\boldsymbol{g}_t(t_f)=0 \tag{21}$$

where $H_{\boldsymbol{x}}$ and $H_{\boldsymbol{u}}$ represent the partial derivatives of $H$. $\boldsymbol{f}_{\boldsymbol{x}}$ and $\boldsymbol{f}_{\boldsymbol{u}}$ are the Jacobi matrixes of $\boldsymbol{f}$. $\varphi_t$, $\varphi_{\boldsymbol{x}}$, $\boldsymbol{g}_t$, and $\boldsymbol{g}_{\boldsymbol{x}}$ are the right-dimensional first-order partial derivatives.

With the VEM, the EPDE and the EDEs derived from the unconstrained functional (17) (regarded as the Lyapunov functional) are [12]

$$\frac{\partial}{\partial\tau}\begin{bmatrix}\boldsymbol{x}(t,\tau)\\\boldsymbol{\lambda}(t,\tau)\\\boldsymbol{u}(t,\tau)\end{bmatrix}=-\boldsymbol{K}\boldsymbol{z},\quad t\in(t_0,t_f) \tag{22}$$

$$\frac{\mathrm{d}}{\mathrm{d}\tau}\begin{bmatrix}\boldsymbol{x}(t_0)\\\boldsymbol{\lambda}(t_0)\\\boldsymbol{u}(t_0)\end{bmatrix}=\boldsymbol{K}\left.\begin{bmatrix}\boldsymbol{W}_{\boldsymbol{x}_0}\left(\boldsymbol{x}_0-\boldsymbol{x}\right)+\left(\dfrac{\partial\boldsymbol{x}}{\partial t}-H_{\boldsymbol{\lambda}}\right)\\\dfrac{\partial\boldsymbol{\lambda}}{\partial t}+H_{\boldsymbol{x}}\\\boldsymbol{z}_{\boldsymbol{u}}\end{bmatrix}\right|_{t_0} \tag{23}$$

$$\frac{\mathrm{d}}{\mathrm{d}\tau}\begin{bmatrix}\boldsymbol{x}(t_f)\\ \boldsymbol{\lambda}(t_f)\\ \boldsymbol{u}(t_f)\end{bmatrix}=-\boldsymbol{K}\left.\begin{bmatrix}\boldsymbol{g_x}^{\mathrm{T}}\boldsymbol{W}_{\boldsymbol{x}_f}\boldsymbol{g}-\left(\varphi_{\boldsymbol{xx}}+\dfrac{\partial(\boldsymbol{g_x}^{\mathrm{T}}\boldsymbol{\pi})}{\partial\boldsymbol{x}}\right)^{\mathrm{T}}\boldsymbol{W}_{\boldsymbol{\lambda}_f}\left(\boldsymbol{\lambda}-\varphi_{\boldsymbol{x}}-\boldsymbol{g_x}^{\mathrm{T}}\boldsymbol{\pi}\right)+w_H(H+\varphi_t+\boldsymbol{\pi}^{\mathrm{T}}\boldsymbol{g}_t)(H_{\boldsymbol{x}}+\varphi_{\boldsymbol{x}t}+\boldsymbol{\pi}^{\mathrm{T}}\boldsymbol{g}_{\boldsymbol{x}t})+\left(\dfrac{\partial\boldsymbol{x}}{\partial t}-H_{\boldsymbol{\lambda}}\right)\\ \boldsymbol{W}_{\boldsymbol{\lambda}_f}\left(\boldsymbol{\lambda}-\varphi_{\boldsymbol{x}}-\boldsymbol{g_x}^{\mathrm{T}}\boldsymbol{\pi}\right)+w_H(H+\varphi_t+\boldsymbol{\pi}^{\mathrm{T}}\boldsymbol{g}_t)\boldsymbol{f}+\left(\dfrac{\partial\boldsymbol{\lambda}}{\partial t}+H_{\boldsymbol{x}}\right)\\ w_H\left(H+\varphi_t+\boldsymbol{\pi}^{\mathrm{T}}\boldsymbol{g}_t\right)H_{\boldsymbol{u}}\end{bmatrix}\right|_{t_f} \tag{24}$$

$$\frac{\mathrm{d}t_f}{\mathrm{d}\tau}=-k_{t_f}h(t_f) \tag{25}$$

$$\frac{\mathrm{d}\boldsymbol{\pi}}{\mathrm{d}\tau}=\boldsymbol{K}_{\boldsymbol{\pi}}\boldsymbol{g_x}(t_f)\boldsymbol{W}_{\boldsymbol{\lambda}_f}\left(\boldsymbol{\lambda}(t_f)-\varphi_{\boldsymbol{x}}(t_f)-\boldsymbol{g_x}^{\mathrm{T}}(t_f)\boldsymbol{\pi}\right) \tag{26}$$

where

$$z(t)=\begin{bmatrix}z_{\boldsymbol{x}}\\ z_{\boldsymbol{\lambda}}\\ z_{\boldsymbol{u}}\end{bmatrix}=H_{\boldsymbol{yy}}\begin{bmatrix}\left(H_{\boldsymbol{x}}+\dfrac{\partial\boldsymbol{\lambda}}{\partial t}\right)\\ \left(\boldsymbol{f}-\dfrac{\partial\boldsymbol{x}}{\partial t}\right)\\ H_{\boldsymbol{u}}\end{bmatrix}-\frac{\partial}{\partial t}\begin{bmatrix}\left(\dfrac{\partial\boldsymbol{x}}{\partial t}-\boldsymbol{f}\right)\\ \left(\dfrac{\partial\boldsymbol{\lambda}}{\partial t}+H_{\boldsymbol{x}}\right)\\ \boldsymbol{0}\end{bmatrix} \tag{27}$$

$$\begin{aligned}h=\;&2\boldsymbol{g}_t^{\mathrm{T}}\boldsymbol{W}_{\boldsymbol{x}_f}\boldsymbol{g}-2\left(\varphi_{\boldsymbol{x}t}+\boldsymbol{g}_{\boldsymbol{x}t}^{\mathrm{T}}\boldsymbol{\pi}\right)^{\mathrm{T}}\boldsymbol{W}_{\boldsymbol{\lambda}_f}\left(\boldsymbol{\lambda}(t_f)-\varphi_{\boldsymbol{x}}(t_f)-\boldsymbol{g_x}^{\mathrm{T}}(t_f)\boldsymbol{\pi}\right)\\ &+2w_H(H+\varphi_t+\boldsymbol{\pi}^{\mathrm{T}}\boldsymbol{g}_t)(H_t+\varphi_{tt}+\boldsymbol{\pi}^{\mathrm{T}}\boldsymbol{g}_{tt})\\ &+H_{\boldsymbol{x}}^{\mathrm{T}}H_{\boldsymbol{x}}+\boldsymbol{f}^{\mathrm{T}}\boldsymbol{f}+H_{\boldsymbol{u}}^{\mathrm{T}}H_{\boldsymbol{u}}-\left(\frac{\partial\boldsymbol{x}}{\partial t}\right)^{\mathrm{T}}\frac{\partial\boldsymbol{x}}{\partial t}-\left(\frac{\partial\boldsymbol{\lambda}}{\partial t}\right)^{\mathrm{T}}\frac{\partial\boldsymbol{\lambda}}{\partial t}\end{aligned} \tag{28}$$

$H_{\boldsymbol{yy}}=\begin{bmatrix}H_{\boldsymbol{xx}} & \boldsymbol{f_x}^{\mathrm{T}} & H_{\boldsymbol{xu}}\\ \boldsymbol{f_x} & \boldsymbol{0} & \boldsymbol{f_u}\\ H_{\boldsymbol{ux}} & \boldsymbol{f_u}^{\mathrm{T}} & H_{\boldsymbol{uu}}\end{bmatrix}$ is the Hessian matrix of $H$, $\boldsymbol{g}_{tt}$ and $\boldsymbol{g}_{\boldsymbol{x}t}$ are second-order partial derivatives of the function $\boldsymbol{g}$, and $\varphi_{\boldsymbol{x}t}$ and $\varphi_{\boldsymbol{xx}}$ are second-order partial derivatives of the function $\varphi$. $\boldsymbol{K}$ is a $(2n+m)\times(2n+m)$ dimensional positive-definite gain matrix, $k_{t_f}$ is a positive gain scalar, and $\boldsymbol{K}_{\boldsymbol{\pi}}$ is a $q\times q$ dimensional positive-definite gain matrix.

**Theorem 1**: Solving the IVP with respect to $\tau$, defined by the evolution equations (22)-(26) with arbitrary initial conditions of $\left.\begin{bmatrix}\boldsymbol{x}(t,\tau)\\ \boldsymbol{\lambda}(t,\tau)\\ \boldsymbol{u}(t,\tau)\end{bmatrix}\right|_{\tau=0}$, $t_f\big|_{\tau=0}$, and $\boldsymbol{\pi}\big|_{\tau=0}$, the solution $\left.\begin{bmatrix}\boldsymbol{x}(t,\tau)\\ \boldsymbol{\lambda}(t,\tau)\\ \boldsymbol{u}(t,\tau)\end{bmatrix}\right|_{\tau=+\infty}$, $t_f\big|_{\tau=+\infty}$, and $\boldsymbol{\pi}\big|_{\tau=+\infty}$ will satisfy the feasibility conditions (13)-(15) and the optimality conditions (18)-(21).

**Proof**: Upon the infinite-dimensional dynamics governed by Eqs. (22)-(26), no matter what initial conditions for $\boldsymbol{x}$, $\boldsymbol{\lambda}$, $\boldsymbol{u}$, $t_f$ and $\boldsymbol{\pi}$ are given at $\tau=0$, we have $\dfrac{\delta\bar{J}}{\delta\tau}\leq 0$ and $\dfrac{\delta\bar{J}}{\delta\tau}=0$ only when $\bar{J}$ reaches its minimum. That is, Eq. (17) is the Lyapunov functional in general. By Lemma 1, we may claim that the minimum solution of the unconstrained functional (17) is an asymptotically stable solution. Thus, $\lim\limits_{\tau\to+\infty}\bar{J}=0$ and the solution will meet the feasibility conditions (13)-(15) and the optimality conditions (18)-(21). ■

**Remark 1**: Since $\boldsymbol{W}_{\boldsymbol{x}_0}$ may be set arbitrarily large in the functional (17) without changing its minimum solution, from a limit viewpoint and with the flexibility in setting the gain matrix, we may modify Eq. (23) as

$$\frac{\mathrm{d}}{\mathrm{d}\tau}\begin{bmatrix}\boldsymbol{x}(t_0)\\ \boldsymbol{\lambda}(t_0)\\ \boldsymbol{u}(t_0)\end{bmatrix}=\boldsymbol{K}\left.\begin{bmatrix}\boldsymbol{x}_0-\boldsymbol{x}\\ \dfrac{\partial\boldsymbol{\lambda}}{\partial t}+H_{\boldsymbol{x}}\\ z_{\boldsymbol{u}}\end{bmatrix}\right|_{t_0} \tag{29}$$

**Remark 2**: Since $\boldsymbol{u}(t_f)$ will reach its equilibrium solution on condition that

$$H_{\boldsymbol{u}}(t_f)=\boldsymbol{0} \tag{30}$$

Eq. (24) may be adapted as

$$\frac{\mathrm{d}}{\mathrm{d}\tau}\begin{bmatrix}\boldsymbol{x}(t_f)\\ \boldsymbol{\lambda}(t_f)\\ \boldsymbol{u}(t_f)\end{bmatrix}=-\boldsymbol{K}\left.\begin{bmatrix}\boldsymbol{g}_{\boldsymbol{x}}{}^{\mathrm{T}}\boldsymbol{W}_{\boldsymbol{x}_f}\boldsymbol{g}-\left(\varphi_{\boldsymbol{xx}}+\dfrac{\partial(\boldsymbol{g}_{\boldsymbol{x}}{}^{\mathrm{T}}\boldsymbol{\pi})}{\partial\boldsymbol{x}}\right)^{\mathrm{T}}\boldsymbol{W}_{\lambda_f}\left(\boldsymbol{\lambda}-\varphi_{\boldsymbol{x}}-\boldsymbol{g}_{\boldsymbol{x}}{}^{\mathrm{T}}\boldsymbol{\pi}\right)+w_H(H+\varphi_t+\boldsymbol{\pi}^{\mathrm{T}}\boldsymbol{g}_t)(H_{\boldsymbol{x}}+\varphi_{\boldsymbol{x}t}+\boldsymbol{\pi}^{\mathrm{T}}\boldsymbol{g}_{\boldsymbol{x}t})+\left(\dfrac{\partial\boldsymbol{x}}{\partial t}-H_{\boldsymbol{\lambda}}\right)\\ \boldsymbol{W}_{\lambda_f}\left(\boldsymbol{\lambda}-\varphi_{\boldsymbol{x}}-\boldsymbol{g}_{\boldsymbol{x}}{}^{\mathrm{T}}\boldsymbol{\pi}\right)+w_H(H+\varphi_t+\boldsymbol{\pi}^{\mathrm{T}}\boldsymbol{g}_t)\boldsymbol{f}+\left(\dfrac{\partial\boldsymbol{\lambda}}{\partial t}+H_{\boldsymbol{x}}\right)\\ z_{\boldsymbol{u}}+\dfrac{\partial\boldsymbol{u}}{\partial t}\dfrac{\mathrm{d}t_f}{\mathrm{d}\tau}\end{bmatrix}\right|_{t_f} \tag{31}$$

which will also preserve the same solution.

**Remark 3**: If in Problem 1 the terminal time $t_f$ is fixed, then the unconstrained foundational is formulated as

$$\begin{aligned}\bar{J}&=\left(\boldsymbol{x}(t_0)-\boldsymbol{x}_0\right)^{\mathrm{T}}\boldsymbol{W}_{\boldsymbol{x}_0}\left(\boldsymbol{x}(t_0)-\boldsymbol{x}_0\right)+\boldsymbol{g}^{\mathrm{T}}\boldsymbol{W}_{\boldsymbol{x}_f}\boldsymbol{g}+\left(\boldsymbol{\lambda}(t_f)-\varphi_{\boldsymbol{x}}(t_f)-\boldsymbol{g}_{\boldsymbol{x}}{}^{\mathrm{T}}(t_f)\boldsymbol{\pi}\right)^{\mathrm{T}}\boldsymbol{W}_{\lambda_f}\left(\boldsymbol{\lambda}(t_f)-\varphi_{\boldsymbol{x}}(t_f)-\boldsymbol{g}_{\boldsymbol{x}}{}^{\mathrm{T}}(t_f)\boldsymbol{\pi}\right)\\ &\quad+\int_{t_0}^{t_f}\left\{\left(\dot{\boldsymbol{x}}-H_{\boldsymbol{\lambda}}\right)^{\mathrm{T}}\left(\dot{\boldsymbol{x}}-H_{\boldsymbol{\lambda}}\right)+\left(\dot{\boldsymbol{\lambda}}+H_{\boldsymbol{x}}\right)^{\mathrm{T}}\left(\dot{\boldsymbol{\lambda}}+H_{\boldsymbol{x}}\right)+H_{\boldsymbol{u}}{}^{\mathrm{T}}H_{\boldsymbol{u}}\right\}\mathrm{d}t\end{aligned} \tag{32}$$

and the resulting evolution equations may be Eqs. (22), (26), (29), and

$$\frac{\mathrm{d}}{\mathrm{d}\tau}\begin{bmatrix}\boldsymbol{x}(t_f)\\ \boldsymbol{\lambda}(t_f)\\ \boldsymbol{u}(t_f)\end{bmatrix}=-\boldsymbol{K}\left.\begin{bmatrix}\boldsymbol{g}_{\boldsymbol{x}}{}^{\mathrm{T}}\boldsymbol{W}_{\boldsymbol{x}_f}\boldsymbol{g}-\left(\varphi_{\boldsymbol{xx}}+\dfrac{\partial(\boldsymbol{g}_{\boldsymbol{x}}{}^{\mathrm{T}}\boldsymbol{\pi})}{\partial\boldsymbol{x}}\right)^{\mathrm{T}}\boldsymbol{W}_{\lambda_f}\left(\boldsymbol{\lambda}-\varphi_{\boldsymbol{x}}-\boldsymbol{g}_{\boldsymbol{x}}{}^{\mathrm{T}}\boldsymbol{\pi}\right)+\left(\dfrac{\partial\boldsymbol{x}}{\partial t}-H_{\boldsymbol{\lambda}}\right)\\ \boldsymbol{W}_{\lambda_f}\left(\boldsymbol{\lambda}-\varphi_{\boldsymbol{x}}-\boldsymbol{g}_{\boldsymbol{x}}{}^{\mathrm{T}}\boldsymbol{\pi}\right)+\left(\dfrac{\partial\boldsymbol{\lambda}}{\partial t}+H_{\boldsymbol{x}}\right)\\ z_{\boldsymbol{u}}\end{bmatrix}\right|_{t_f} \tag{33}$$

which is consistent with Eq. (31).

**Remark 4**: If in Problem 1 the terminal states are free, i.e., $\boldsymbol{g}$ vanishes, then the unconstrained foundational is simplified as

$$\begin{aligned}\bar{J}&=\left(\boldsymbol{x}(t_0)-\boldsymbol{x}_0\right)^{\mathrm{T}}\boldsymbol{W}_{\boldsymbol{x}_0}\left(\boldsymbol{x}(t_0)-\boldsymbol{x}_0\right)+\left(\boldsymbol{\lambda}(t_f)-\varphi_{\boldsymbol{x}}(t_f)\right)^{\mathrm{T}}\boldsymbol{W}_{\lambda_f}\left(\boldsymbol{\lambda}(t_f)-\varphi_{\boldsymbol{x}}(t_f)\right)\\ &\quad+w_H\left(H(t_f)+\varphi_t(t_f)\right)^2+\int_{t_0}^{t_f}\left\{\left(\dot{\boldsymbol{x}}-H_{\boldsymbol{\lambda}}\right)^{\mathrm{T}}\left(\dot{\boldsymbol{x}}-H_{\boldsymbol{\lambda}}\right)+\left(\dot{\boldsymbol{\lambda}}+H_{\boldsymbol{x}}\right)^{\mathrm{T}}\left(\dot{\boldsymbol{\lambda}}+H_{\boldsymbol{x}}\right)+H_{\boldsymbol{u}}{}^{\mathrm{T}}H_{\boldsymbol{u}}\right\}\mathrm{d}t\end{aligned} \tag{34}$$

and the resulting evolution equations developed may be Eqs. (22), (25), (29), and (31) with terms regarding $\boldsymbol{g}$ removed.

Employing the primary form of the first evolution equation to seek the optimal solution, the anticipated variable evolving along the variation time $\tau$, as depicted in Fig. 1, includes the control variables, the state variables and the costate variables. When we apply the semi-discrete method to solve the EPDE (22), the state, costate, and control variables all need to be discretized along the normal time dimension $t$.

*C. The compact formulation*

With the similar treatment in establishing the third evolution equation [20], we do not solve for the states and the costates from the variation motion while they are computed in terms of the state equation (13) and the costate equation (18) with the boundary conditions given by Eqs.(14) and (20), namely

$$\boldsymbol{x}(t) = \boldsymbol{x}_0 + \int_{t_0}^{t} \boldsymbol{f}\left(\boldsymbol{x}, \boldsymbol{u}, s\right) \mathrm{d}s \tag{35}$$

$$\boldsymbol{\lambda}(t) = \varphi_{\boldsymbol{x}}(t_f) + \boldsymbol{g}_{\boldsymbol{x}}^{\mathrm{T}}(t_f)\boldsymbol{\pi} + \int_{t}^{t_f} \left(L_{\boldsymbol{x}} + \boldsymbol{f}_{\boldsymbol{x}}^{\mathrm{T}}\boldsymbol{\lambda}\right) \mathrm{d}s \tag{36}$$

In particular, we have derived the explicit expression of $\boldsymbol{\lambda}(t)$ for Eq. (36) in Ref. [15], that is

$$\boldsymbol{\lambda}(t) = \varphi_{\boldsymbol{x}}(t) + \boldsymbol{\Phi}_o^{\mathrm{T}}(t_f, t)\boldsymbol{g}_{\boldsymbol{x}}^{\mathrm{T}}(t_f)\boldsymbol{\pi} + \int_{t}^{t_f} \boldsymbol{\Phi}_o^{\mathrm{T}}(\sigma, t)\overline{L}_{\boldsymbol{x}}(\sigma)\,\mathrm{d}\sigma \tag{37}$$

where

$$\overline{L} = \varphi_t + \varphi_{\boldsymbol{x}}^{\mathrm{T}}\boldsymbol{f} + L \tag{38}$$

$\boldsymbol{\Phi}_o(\sigma, t)$ is the $n \times n$ dimensional state transition matrix from time point $t$ to time point $\sigma$, and it satisfies

$$\frac{\partial \boldsymbol{\Phi}_o(\sigma, t)}{\partial t} = -\boldsymbol{\Phi}_o(\sigma, t)\boldsymbol{f}_{\boldsymbol{x}}(t) \tag{39}$$

In this way, we consider Problem 1 within the quasi-feasible solution domain $\mathbb{D}_q$, in which the solutions including variables $\boldsymbol{x}$, $\boldsymbol{\lambda}$, $\boldsymbol{u}$ and parameters $\boldsymbol{\pi}$ satisfy Eqs. (35) and (37). Now the unconstrained functional are constructed as

$$\overline{J} = \boldsymbol{g}^{\mathrm{T}}\mathbf{W}_{\boldsymbol{x}_f}\boldsymbol{g} + w_H\left(H(t_f) + \varphi_t(t_f) + \boldsymbol{\pi}^{\mathrm{T}}\boldsymbol{g}_t(t_f)\right)^2 + \int_{t_0}^{t_f} H_{\boldsymbol{u}}^{\mathrm{T}} H_{\boldsymbol{u}}\mathrm{d}t \tag{40}$$

where $\boldsymbol{W}_{\boldsymbol{x}_f}$ and $w_H$ are the weights.

Differentiating Eq. (40) with respect to the variation time $\tau$ gives

$$\begin{aligned}
\frac{\delta \overline{J}}{\delta \tau} = {} & 2\boldsymbol{g}^{\mathrm{T}}\mathbf{W}_{\boldsymbol{x}_f}\left(\boldsymbol{g}_{\boldsymbol{x}}(t_f)\frac{\delta \boldsymbol{x}(t_f)}{\delta \tau} + (\boldsymbol{g}_{\boldsymbol{x}}\boldsymbol{f} + \boldsymbol{g}_t)\Big|_{t_f}\frac{\delta t_f}{\delta \tau}\right) \\
& + 2w_H(H + \varphi_t + \boldsymbol{\pi}^{\mathrm{T}}\boldsymbol{g}_t)(H_{\boldsymbol{x}} + \varphi_{\boldsymbol{x}t} + \boldsymbol{\pi}^{\mathrm{T}}\boldsymbol{g}_{\boldsymbol{x}t})^{\mathrm{T}}\Big|_{t_f}\frac{\delta \boldsymbol{x}(t_f)}{\delta \tau} + 2w_H(H + \varphi_t + \boldsymbol{\pi}^{\mathrm{T}}\boldsymbol{g}_t)\boldsymbol{f}^{\mathrm{T}}\Big|_{t_f}\frac{\delta \boldsymbol{\lambda}(t_f)}{\delta \tau} \\
& + 2w_H(H + \varphi_t + \boldsymbol{\pi}^{\mathrm{T}}\boldsymbol{g}_t)\left\{(\varphi_{\boldsymbol{x}t} + \boldsymbol{\pi}^{\mathrm{T}}\boldsymbol{g}_{\boldsymbol{x}t})^{\mathrm{T}}\boldsymbol{f} + (H_t + \varphi_{tt} + \boldsymbol{\pi}^{\mathrm{T}}\boldsymbol{g}_{tt})\right\}\Big|_{t_f}\frac{\delta t_f}{\delta \tau} \\
& + 2w_H(H + \varphi_t + \boldsymbol{\pi}^{\mathrm{T}}\boldsymbol{g}_t)\boldsymbol{g}_t^{\mathrm{T}}\Big|_{t_f}\frac{\delta \boldsymbol{\pi}}{\delta \tau} \\
& + 2\int_{t_0}^{t_f}\left(H_{\boldsymbol{u}}^{\mathrm{T}}H_{\boldsymbol{uu}}\frac{\delta \boldsymbol{u}}{\delta \tau} + H_{\boldsymbol{u}}^{\mathrm{T}}H_{\boldsymbol{ux}}\frac{\delta \boldsymbol{x}}{\delta \tau} + H_{\boldsymbol{u}}^{\mathrm{T}}\boldsymbol{f}_{\boldsymbol{u}}^{\mathrm{T}}\frac{\delta \boldsymbol{\lambda}}{\delta \tau}\right)\mathrm{d}t + (H_{\boldsymbol{u}}^{\mathrm{T}}H_{\boldsymbol{u}})\Big|_{t_f}\frac{\delta t_f}{\delta \tau}
\end{aligned} \tag{41}$$

From the state equation (13), we derived the explicit variation evolution relation between $\dfrac{\delta \boldsymbol{x}}{\delta \tau}$ and $\dfrac{\delta \boldsymbol{u}}{\delta \tau}$ as [14]

$$\frac{\delta \boldsymbol{x}}{\delta \tau} = \int_{t_0}^{t} \boldsymbol{\Phi}_o(t, s)\boldsymbol{f}_{\boldsymbol{u}}(s)\frac{\delta \boldsymbol{u}}{\delta \tau}(s)\,\mathrm{d}s \tag{42}$$

At the terminal time $t_f$, we have

$$\frac{\delta \boldsymbol{x}(t_f)}{\delta \tau}=\int_{t_0}^{t_f}\boldsymbol{\Phi}_o(t_f,s)\boldsymbol{f}_{\boldsymbol{u}}(s)\frac{\delta \boldsymbol{u}}{\delta \tau}(s)\,\mathrm{d}\,s \tag{43}$$

From the costate expression (37), the variation evolution relation may be established as

$$\begin{aligned}\frac{\delta \boldsymbol{\lambda}}{\delta \tau}=&\,\varphi_{\boldsymbol{xx}}(t)\frac{\delta \boldsymbol{x}}{\delta \tau}+\boldsymbol{\Phi}_o{}^{\mathrm{T}}(t_f,t)\left(\boldsymbol{g}_{\boldsymbol{xx}}{}^{\mathrm{T}}(t_f)\boldsymbol{\pi}\right)\frac{\delta \boldsymbol{x}(t_f)}{\delta \tau}+\boldsymbol{\Phi}_o{}^{\mathrm{T}}(t_f,t)\boldsymbol{g}_{\boldsymbol{x}}{}^{\mathrm{T}}(t_f)\frac{\delta \boldsymbol{\pi}}{\delta \tau}\\&+\boldsymbol{\Phi}_o{}^{\mathrm{T}}(t_f,t)\left\{\left(\boldsymbol{g}_{\boldsymbol{xx}}{}^{\mathrm{T}}(t_f)\boldsymbol{\pi}\right)\boldsymbol{f}(t_f)+\boldsymbol{g}_{\boldsymbol{x}t}{}^{\mathrm{T}}(t_f)\boldsymbol{\pi}+\boldsymbol{f}_{\boldsymbol{x}}{}^{\mathrm{T}}(t_f)\boldsymbol{g}_{\boldsymbol{x}}{}^{\mathrm{T}}(t_f)\boldsymbol{\pi}+\bar{L}_{\boldsymbol{x}}(t_f)\right\}\frac{\delta t_f}{\delta \tau}\\&+\int_{t}^{t_f}\boldsymbol{\Phi}_o{}^{\mathrm{T}}(\sigma,t)\left(\bar{L}_{\boldsymbol{xx}}(\sigma)\frac{\delta \boldsymbol{x}}{\delta \tau}+\bar{L}_{\boldsymbol{x}u}(\sigma)\frac{\delta \boldsymbol{u}}{\delta \tau}\right)\mathrm{d}\,\sigma\end{aligned} \tag{44}$$

where it is defined that $\boldsymbol{g}_{\boldsymbol{xx}}{}^{\mathrm{T}}(t_f)\boldsymbol{\pi}=\left.\frac{\partial(\boldsymbol{g}_{\boldsymbol{x}}{}^{\mathrm{T}}\boldsymbol{\pi})}{\partial \boldsymbol{x}}\right|_{t_f}$ . Correspondingly, there is

$$\frac{\delta \boldsymbol{\lambda}(t_f)}{\delta \tau}=\left.\left(\varphi_{\boldsymbol{xx}}+\frac{\partial(\boldsymbol{g}_{\boldsymbol{x}}{}^{\mathrm{T}}\boldsymbol{\pi})}{\partial \boldsymbol{x}}\right)\right|_{t_f}\frac{\delta \boldsymbol{x}(t_f)}{\delta \tau}+\boldsymbol{g}_{\boldsymbol{x}}{}^{\mathrm{T}}(t_f)\frac{\delta \boldsymbol{\pi}}{\delta \tau}+\left.\left(\frac{\partial(\boldsymbol{g}_{\boldsymbol{x}}{}^{\mathrm{T}}\boldsymbol{\pi})}{\partial \boldsymbol{x}}\boldsymbol{f}+\boldsymbol{g}_{\boldsymbol{x}t}{}^{\mathrm{T}}\boldsymbol{\pi}+\boldsymbol{f}_{\boldsymbol{x}}{}^{\mathrm{T}}\boldsymbol{g}_{\boldsymbol{x}}{}^{\mathrm{T}}\boldsymbol{\pi}+\bar{L}_{\boldsymbol{x}}\right)\right|_{t_f}\frac{\delta t_f}{\delta \tau} \tag{45}$$

Substituting Eq. (45) into Eq. (41) gives

$$\begin{aligned}\frac{\delta \bar{J}}{\delta \tau}=&\,2\left.\left(\boldsymbol{g}^{\mathrm{T}}\boldsymbol{W}_{\boldsymbol{x}_f}\boldsymbol{g}_{\boldsymbol{x}}+w_H(H+\varphi_t+\boldsymbol{\pi}^{\mathrm{T}}\boldsymbol{g}_t)\left\{H_{\boldsymbol{x}}+\varphi_{\boldsymbol{x}t}+\boldsymbol{\pi}^{\mathrm{T}}\boldsymbol{g}_{\boldsymbol{x}t}+\left(\varphi_{\boldsymbol{xx}}+\frac{\partial(\boldsymbol{g}_{\boldsymbol{x}}{}^{\mathrm{T}}\boldsymbol{\pi})}{\partial \boldsymbol{x}}\right)^{\mathrm{T}}\boldsymbol{f}\right\}^{\mathrm{T}}\right)\right|_{t_f}\frac{\delta \boldsymbol{x}(t_f)}{\delta \tau}\\&+2\left.\left(\boldsymbol{g}^{\mathrm{T}}\boldsymbol{W}_{\boldsymbol{x}_f}(\boldsymbol{g}_{\boldsymbol{x}}\boldsymbol{f}+\boldsymbol{g}_t)+\frac{1}{2}H_{\boldsymbol{u}}{}^{\mathrm{T}}H_{\boldsymbol{u}}+w_H(H+\varphi_t+\boldsymbol{\pi}^{\mathrm{T}}\boldsymbol{g}_t)\left\{\begin{array}{l}\boldsymbol{f}^{\mathrm{T}}(\varphi_{\boldsymbol{x}t}+2\boldsymbol{\pi}^{\mathrm{T}}\boldsymbol{g}_{\boldsymbol{x}t}+\frac{\partial(\boldsymbol{g}_{\boldsymbol{x}}{}^{\mathrm{T}}\boldsymbol{\pi})}{\partial \boldsymbol{x}}\boldsymbol{f}+\boldsymbol{f}_{\boldsymbol{x}}{}^{\mathrm{T}}\boldsymbol{g}_{\boldsymbol{x}}{}^{\mathrm{T}}\boldsymbol{\pi}+\bar{L}_{\boldsymbol{x}})\\+(H_t+\varphi_{tt}+\boldsymbol{\pi}^{\mathrm{T}}\boldsymbol{g}_{tt})\end{array}\right\}\right)\right|_{t_f}\frac{\delta t_f}{\delta \tau}\\&+2\left.\left(w_H(H+\varphi_t+\boldsymbol{\pi}^{\mathrm{T}}\boldsymbol{g}_t)(\boldsymbol{g}_t{}^{\mathrm{T}}+\boldsymbol{f}^{\mathrm{T}}\boldsymbol{g}_{\boldsymbol{x}}{}^{\mathrm{T}})\right)\right|_{t_f}\frac{\delta \boldsymbol{\pi}}{\delta \tau}\\&+2\int_{t_0}^{t_f}\left(H_{\boldsymbol{u}}{}^{\mathrm{T}}H_{\boldsymbol{uu}}\frac{\delta \boldsymbol{u}}{\delta \tau}\right)\mathrm{d}t+2\int_{t_0}^{t_f}\left(H_{\boldsymbol{u}}{}^{\mathrm{T}}H_{\boldsymbol{ux}}\frac{\delta \boldsymbol{x}}{\delta \tau}\right)\mathrm{d}t+2\int_{t_0}^{t_f}\left(H_{\boldsymbol{u}}{}^{\mathrm{T}}\boldsymbol{f}_{\boldsymbol{u}}{}^{\mathrm{T}}\frac{\delta \boldsymbol{\lambda}}{\delta \tau}\right)\mathrm{d}t\end{aligned} \tag{46}$$

Using the relations given by Eqs. (42), (44) and exchanging the order in the multiple integral, we may further derive that

$$\int_{t_0}^{t_f}\left(H_{\boldsymbol{u}}{}^{\mathrm{T}}H_{\boldsymbol{ux}}\frac{\delta \boldsymbol{x}}{\delta \tau}\right)\mathrm{d}t=\int_{t_0}^{t_f}\left(\int_{s}^{t_f}H_{\boldsymbol{u}}{}^{\mathrm{T}}H_{\boldsymbol{ux}}\boldsymbol{\Phi}_o(t,s)\mathrm{d}t\right)\boldsymbol{f}_{\boldsymbol{u}}(s)\frac{\delta \boldsymbol{u}}{\delta \tau}(s)\,\mathrm{d}\,s \tag{47}$$

and

$$\begin{aligned}\int_{t_0}^{t_f}\left(H_{\boldsymbol{u}}{}^{\mathrm{T}}\boldsymbol{f}_{\boldsymbol{u}}{}^{\mathrm{T}}\frac{\delta \boldsymbol{\lambda}}{\delta \tau}\right)\mathrm{d}t=&\int_{t_0}^{t_f}\left(\int_{s}^{t_f}H_{\boldsymbol{u}}{}^{\mathrm{T}}\boldsymbol{f}_{\boldsymbol{u}}{}^{\mathrm{T}}\varphi_{\boldsymbol{xx}}\boldsymbol{\Phi}_o(t,s)\mathrm{d}t\right)\boldsymbol{f}_{\boldsymbol{u}}(s)\frac{\delta \boldsymbol{u}}{\delta \tau}(s)\,\mathrm{d}\,s\\&+\int_{t_0}^{t_f}\left(\int_{s}^{t_f}\left(\int_{t_0}^{\sigma}H_{\boldsymbol{u}}{}^{\mathrm{T}}\boldsymbol{f}_{\boldsymbol{u}}{}^{\mathrm{T}}\boldsymbol{\Phi}_o{}^{\mathrm{T}}(\sigma,t)\mathrm{d}t\right)\bar{L}_{\boldsymbol{xx}}(\sigma)\boldsymbol{\Phi}_o(\sigma,s)\,\mathrm{d}\,\sigma\right)\boldsymbol{f}_{\boldsymbol{u}}(s)\frac{\delta \boldsymbol{u}}{\delta \tau}(s)\,\mathrm{d}\,s\\&+\int_{t_0}^{t_f}\left(\int_{t_0}^{s}H_{\boldsymbol{u}}{}^{\mathrm{T}}\boldsymbol{f}_{\boldsymbol{u}}{}^{\mathrm{T}}\boldsymbol{\Phi}_o{}^{\mathrm{T}}(s,t)\mathrm{d}t\right)\bar{L}_{\boldsymbol{x}u}(s)\frac{\delta \boldsymbol{u}}{\delta \tau}(s)\,\mathrm{d}\,s\\&+\int_{t_0}^{t_f}\left(\int_{t_0}^{t_f}H_{\boldsymbol{u}}{}^{\mathrm{T}}\boldsymbol{f}_{\boldsymbol{u}}{}^{\mathrm{T}}\boldsymbol{\Phi}_o{}^{\mathrm{T}}(t_f,t)\mathrm{d}t\right)\left(\boldsymbol{g}_{\boldsymbol{xx}}{}^{\mathrm{T}}(t_f)\boldsymbol{\pi}\right)\boldsymbol{\Phi}_o(t_f,s)\boldsymbol{f}_{\boldsymbol{u}}(s)\frac{\delta \boldsymbol{u}}{\delta \tau}(s)\,\mathrm{d}\,s\\&+\left(\int_{t_0}^{t_f}H_{\boldsymbol{u}}{}^{\mathrm{T}}\boldsymbol{f}_{\boldsymbol{u}}{}^{\mathrm{T}}\boldsymbol{\Phi}_o{}^{\mathrm{T}}(t_f,t)\mathrm{d}t\right)\boldsymbol{g}_{\boldsymbol{x}}{}^{\mathrm{T}}(t_f)\frac{\delta \boldsymbol{\pi}}{\delta \tau}\\&+\left(\int_{t_0}^{t_f}H_{\boldsymbol{u}}{}^{\mathrm{T}}\boldsymbol{f}_{\boldsymbol{u}}{}^{\mathrm{T}}\boldsymbol{\Phi}_o{}^{\mathrm{T}}(t_f,t)\mathrm{d}t\right)\left\{\left(\boldsymbol{g}_{\boldsymbol{xx}}{}^{\mathrm{T}}(t_f)\boldsymbol{\pi}\right)\boldsymbol{f}(t_f)+\boldsymbol{g}_{\boldsymbol{x}t}{}^{\mathrm{T}}(t_f)\boldsymbol{\pi}+\boldsymbol{f}_{\boldsymbol{x}}{}^{\mathrm{T}}(t_f)\boldsymbol{g}_{\boldsymbol{x}}{}^{\mathrm{T}}(t_f)\boldsymbol{\pi}+\bar{L}_{\boldsymbol{x}}(t_f)\right\}\frac{\delta t_f}{\delta \tau}\end{aligned} \tag{48}$$

With Eqs. (43), (47), (48) and define

$$
\begin{aligned}
\boldsymbol{n}_{\boldsymbol{u}}(t) = {} & H_{\boldsymbol{uu}} H_{\boldsymbol{u}} + \bar{L}_{\boldsymbol{x}u}{}^{\mathrm{T}}(t) \left( \int_{t_0}^{t} \boldsymbol{\Phi}_o(t,s) \boldsymbol{f}_{\boldsymbol{u}} H_{\boldsymbol{u}} \mathrm{d}s \right) \\
& + \boldsymbol{f}_{\boldsymbol{u}}{}^{\mathrm{T}}(t) \left( \int_{t}^{t_f} \boldsymbol{\Phi}_o{}^{\mathrm{T}}(\sigma,t) \left\{ H_{\boldsymbol{ux}}{}^{\mathrm{T}} H_{\boldsymbol{u}} + \varphi_{\boldsymbol{xx}} \boldsymbol{f}_{\boldsymbol{u}} H_{\boldsymbol{u}} + \bar{L}_{\boldsymbol{xx}}{}^{\mathrm{T}}(\sigma) \left( \int_{t_0}^{\sigma} \boldsymbol{\Phi}_o(\sigma,s) \boldsymbol{f}_{\boldsymbol{u}} H_{\boldsymbol{u}} \mathrm{d}s \right) \right\} \mathrm{d}\sigma \right) \\
& + \boldsymbol{f}_{\boldsymbol{u}}{}^{\mathrm{T}}(t) \boldsymbol{\Phi}_o{}^{\mathrm{T}}(t_f,t) \left. \left( \begin{array}{l} \boldsymbol{g}_{\boldsymbol{x}}{}^{\mathrm{T}} \boldsymbol{W}_{\boldsymbol{x}_f} \boldsymbol{g} + \left( \dfrac{\partial (\boldsymbol{g}_{\boldsymbol{x}}{}^{\mathrm{T}} \boldsymbol{\pi})}{\partial \boldsymbol{x}} \right)^{\mathrm{T}} \displaystyle\int_{t_0}^{t_f} \boldsymbol{\Phi}_o(t_f,s) \boldsymbol{f}_{\boldsymbol{u}} H_{\boldsymbol{u}} \mathrm{d}s \\ + w_H (H + \varphi_t + \boldsymbol{\pi}^{\mathrm{T}} \boldsymbol{g}_t) \left\{ H_{\boldsymbol{x}} + \varphi_{\boldsymbol{x}t} + \boldsymbol{\pi}^{\mathrm{T}} \boldsymbol{g}_{\boldsymbol{x}t} + \left( \varphi_{\boldsymbol{xx}} + \dfrac{\partial (\boldsymbol{g}_{\boldsymbol{x}}{}^{\mathrm{T}} \boldsymbol{\pi})}{\partial \boldsymbol{x}} \right)^{\mathrm{T}} \boldsymbol{f} \right\} \end{array} \right) \right|_{t_f}
\end{aligned}
\tag{49}
$$

$$
\boldsymbol{n}_{t_f} = \left. \left( \begin{array}{l} \boldsymbol{g}^{\mathrm{T}} \boldsymbol{W}_{\boldsymbol{x}_f} (\boldsymbol{g}_{\boldsymbol{x}} \boldsymbol{f} + \boldsymbol{g}_t) + \dfrac{1}{2} H_{\boldsymbol{u}}{}^{\mathrm{T}} H_{\boldsymbol{u}} + \left( \displaystyle\int_{t_0}^{t_f} \boldsymbol{\Phi}_o(t_f,s) \boldsymbol{f}_{\boldsymbol{u}} H_{\boldsymbol{u}} \mathrm{d}s \right)^{\mathrm{T}} \left( \dfrac{\partial (\boldsymbol{g}_{\boldsymbol{x}}{}^{\mathrm{T}} \boldsymbol{\pi})}{\partial \boldsymbol{x}} \boldsymbol{f} + \boldsymbol{g}_{\boldsymbol{x}t}{}^{\mathrm{T}} \boldsymbol{\pi} + \boldsymbol{f}_{\boldsymbol{x}}{}^{\mathrm{T}} \boldsymbol{g}_{\boldsymbol{x}}{}^{\mathrm{T}} \boldsymbol{\pi} + \bar{L}_{\boldsymbol{x}} \right) \\ + w_H (H + \varphi_t + \boldsymbol{\pi}^{\mathrm{T}} \boldsymbol{g}_t) \left\{ \boldsymbol{f}^{\mathrm{T}} \left( \varphi_{\boldsymbol{x}t} + 2\boldsymbol{\pi}^{\mathrm{T}} \boldsymbol{g}_{\boldsymbol{x}t} + \dfrac{\partial (\boldsymbol{g}_{\boldsymbol{x}}{}^{\mathrm{T}} \boldsymbol{\pi})}{\partial \boldsymbol{x}} \boldsymbol{f} + \boldsymbol{f}_{\boldsymbol{x}}{}^{\mathrm{T}} \boldsymbol{g}_{\boldsymbol{x}}{}^{\mathrm{T}} \boldsymbol{\pi} + \bar{L}_{\boldsymbol{x}} \right) + (H_t + \varphi_{tt} + \boldsymbol{\pi}^{\mathrm{T}} \boldsymbol{g}_{tt}) \right\} \end{array} \right) \right|_{t_f}
\tag{50}
$$

$$
\boldsymbol{n}_{\boldsymbol{\pi}} = w_H (H + \varphi_t + \boldsymbol{\pi}^{\mathrm{T}} \boldsymbol{g}_t)(\boldsymbol{g}_t + \boldsymbol{g}_{\boldsymbol{x}} \boldsymbol{f}) \Big|_{t_f} + \boldsymbol{g}_{\boldsymbol{x}}(t_f) \int_{t_0}^{t_f} \boldsymbol{\Phi}_o(t_f,s) \boldsymbol{f}_{\boldsymbol{u}} H_{\boldsymbol{u}} \mathrm{d}s
\tag{51}
$$

we have

$$
\frac{\delta \bar{J}}{\delta \tau} = 2 \int_{t_0}^{t_f} \left( \boldsymbol{n}_{\boldsymbol{u}}{}^{\mathrm{T}} \frac{\delta \boldsymbol{u}}{\delta \tau} \right) \mathrm{d}t + 2 \boldsymbol{n}_{t_f} \frac{\delta t_f}{\delta \tau} + 2 \boldsymbol{n}_{\boldsymbol{\pi}} \frac{\delta \boldsymbol{\pi}}{\delta \tau}
\tag{52}
$$

To achieve $\dfrac{\delta \bar{J}}{\delta \tau} \leq 0$, we may set the variation dynamic evolution equation for the control $\boldsymbol{u}$ as

$$
\frac{\delta \boldsymbol{u}}{\delta \tau} = -\boldsymbol{K} \boldsymbol{n}_{\boldsymbol{u}}
\tag{53}
$$

for the terminal time $t_f$ as

$$
\frac{\delta t_f}{\delta \tau} = -k_{t_f} \boldsymbol{n}_{t_f}
\tag{54}
$$

and for the Lagrange multiplier parameters $\boldsymbol{\pi}$ as

$$
\frac{\delta \boldsymbol{\pi}}{\delta \tau} = -\boldsymbol{K}_{\boldsymbol{\pi}} \boldsymbol{n}_{\boldsymbol{\pi}}
\tag{55}
$$

where $\boldsymbol{K}$ is a $m \times m$ dimensional positive-definite gain matrix, $k_{t_f}$ is a positive gain scalar, and $\boldsymbol{K}_{\boldsymbol{\pi}}$ is a $q \times q$ dimensional positive-definite gain matrix.

Use the partial differential operator "$\partial$" and the differential operator "$\mathrm{d}$" to reformulate the variation dynamic evolution equations (53)-(55), we may get the following EPDE and EDEs, i.e., the compact form of the first evolution equation, as

$$
\frac{\partial \boldsymbol{u}(t,\tau)}{\partial \tau} = -\boldsymbol{K} \boldsymbol{n}_{\boldsymbol{u}}(t)
\tag{56}
$$

$$
\frac{\mathrm{d} t_f}{\mathrm{d} \tau} = -k_{t_f} \boldsymbol{n}_{t_f}
\tag{57}
$$

$$
\frac{\mathrm{d} \boldsymbol{\pi}}{\mathrm{d} \tau} = -\boldsymbol{K}_{\boldsymbol{\pi}} \boldsymbol{n}_{\boldsymbol{\pi}}
\tag{58}
$$

**Theorem 2**: Solving the IVP with respect to $\tau$, defined by the evolution equations (56)-(58) with $\boldsymbol{x}$ and $\boldsymbol{\lambda}$ determined by Eqs. (35) and (36), from arbitrary initial conditions of $\boldsymbol{u}(t,\tau)\big|_{\tau=0}$, $t_f\big|_{\tau=0}$, and $\boldsymbol{\pi}\big|_{\tau=0}$, when $\tau \to +\infty$, $(\boldsymbol{x},\boldsymbol{\lambda},\boldsymbol{u})$ and $\boldsymbol{\pi}$ will satisfy the feasibility conditions (13)-(15) and the optimality conditions (18)-(21).

**Proof**: With $\boldsymbol{x}$ and $\boldsymbol{\lambda}$ determined by Eqs. (35) and (36), the solution is guaranteed in the quasi-feasible solution domain $\bar{\mathbb{D}}_q$. Thus, the feasibility conditions (13), (14) and the optimality conditions (18), (20) are always met. For the infinite-dimensional dynamics governed by Eqs. (56)-(58), Eq. (40) is the Lyapunov functional in $\bar{\mathbb{D}}_q$ because $\frac{\delta \bar{J}}{\delta \tau} \leq 0$. By Lemma 1, the minimum solution of the functional (40) is an asymptotically stable solution within $\bar{\mathbb{D}}_q$. Thus, when $\tau \to +\infty$, we have $\bar{J}$ converge to zero, which means $(\boldsymbol{x},\boldsymbol{\lambda},\boldsymbol{u})$ and $\boldsymbol{\pi}$ will satisfy the feasibility conditions (15) and the optimality conditions (19), (21). ■

**Remark 5**: If in Problem 1 the terminal time $t_f$ is fixed, then the unconstrained foundational is simplified as

$$\bar{J} = \boldsymbol{g}^{\mathrm{T}} \boldsymbol{W}_{\boldsymbol{x}_f} \boldsymbol{g} + \int_{t_0}^{t_f} H_{\boldsymbol{u}}{}^{\mathrm{T}} H_{\boldsymbol{u}} \mathrm{d}t \tag{59}$$

The resulting EPDE (56) on $\boldsymbol{u}$ is modified with $\boldsymbol{n}_{\boldsymbol{u}}(t)$ being

$$\begin{aligned} \boldsymbol{n}_{\boldsymbol{u}}(t) = {} & H_{\boldsymbol{uu}} H_{\boldsymbol{u}} + \bar{L}_{\boldsymbol{x}u}{}^{\mathrm{T}}(t) \left( \int_{t_0}^{t} \boldsymbol{\Phi}_o(t,s) \boldsymbol{f}_{\boldsymbol{u}} H_{\boldsymbol{u}} \mathrm{d}s \right) \\ & + \boldsymbol{f}_{\boldsymbol{u}}{}^{\mathrm{T}}(t) \left( \int_{t}^{t_f} \boldsymbol{\Phi}_o{}^{\mathrm{T}}(\sigma,t) \left\{ H_{\boldsymbol{ux}}{}^{\mathrm{T}} H_{\boldsymbol{u}} + \varphi_{\boldsymbol{xx}} \boldsymbol{f}_{\boldsymbol{u}} H_{\boldsymbol{u}} + \bar{L}_{\boldsymbol{xx}}{}^{\mathrm{T}}(\sigma) \left( \int_{t_0}^{\sigma} \boldsymbol{\Phi}_o(\sigma,s) \boldsymbol{f}_{\boldsymbol{u}} H_{\boldsymbol{u}} \mathrm{d}s \right) \right\} \mathrm{d}\sigma \right) \\ & + \boldsymbol{f}_{\boldsymbol{u}}{}^{\mathrm{T}}(t) \boldsymbol{\Phi}_o{}^{\mathrm{T}}(t_f,t) \left( \boldsymbol{g}_{\boldsymbol{x}}{}^{\mathrm{T}} \boldsymbol{W}_{\boldsymbol{x}_f} \boldsymbol{g} + \left( \frac{\partial (\boldsymbol{g}_{\boldsymbol{x}}{}^{\mathrm{T}} \boldsymbol{\pi})}{\partial \boldsymbol{x}} \right)^{\mathrm{T}} \int_{t_0}^{t_f} \boldsymbol{\Phi}_o(t_f,s) \boldsymbol{f}_{\boldsymbol{u}} H_{\boldsymbol{u}} \mathrm{d}s \right) \Bigg|_{t_f} \end{aligned} \tag{60}$$

and the EDE (58) on $\boldsymbol{\pi}$ is modified with $\boldsymbol{n}_{\boldsymbol{\pi}}$ being

$$\boldsymbol{n}_{\boldsymbol{\pi}} = \boldsymbol{g}_{\boldsymbol{x}}(t_f) \int_{t_0}^{t_f} \boldsymbol{\Phi}_o(t_f,s) \boldsymbol{f}_{\boldsymbol{u}} H_{\boldsymbol{u}} \mathrm{d}s \tag{61}$$

**Remark 6**: If in Problem 1 the terminal states are free, i.e., $\boldsymbol{g}$ vanishes, then the unconstrained foundational is constructed as

$$\bar{J} = w_H \left( H(t_f) + \varphi_t(t_f) \right)^2 + \int_{t_0}^{t_f} H_{\boldsymbol{u}}{}^{\mathrm{T}} H_{\boldsymbol{u}} \mathrm{d}t \tag{62}$$

Now $\boldsymbol{\pi} = \boldsymbol{0}$ and the EPDE (56) on $\boldsymbol{u}$ is modified with $\boldsymbol{n}_{\boldsymbol{u}}(t)$ being

$$\begin{aligned} \boldsymbol{n}_{\boldsymbol{u}}(t) = {} & H_{\boldsymbol{uu}} H_{\boldsymbol{u}} + \bar{L}_{\boldsymbol{x}u}{}^{\mathrm{T}}(t) \left( \int_{t_0}^{t} \boldsymbol{\Phi}_o(t,s) \boldsymbol{f}_{\boldsymbol{u}} H_{\boldsymbol{u}} \mathrm{d}s \right) \\ & + \boldsymbol{f}_{\boldsymbol{u}}{}^{\mathrm{T}}(t) \left( \int_{t}^{t_f} \boldsymbol{\Phi}_o{}^{\mathrm{T}}(\sigma,t) \left\{ H_{\boldsymbol{ux}}{}^{\mathrm{T}} H_{\boldsymbol{u}} + \varphi_{\boldsymbol{xx}} \boldsymbol{f}_{\boldsymbol{u}} H_{\boldsymbol{u}} + \bar{L}_{\boldsymbol{xx}}{}^{\mathrm{T}}(\sigma) \left( \int_{t_0}^{\sigma} \boldsymbol{\Phi}_o(\sigma,s) \boldsymbol{f}_{\boldsymbol{u}} H_{\boldsymbol{u}} \mathrm{d}s \right) \right\} \mathrm{d}\sigma \right) \\ & + \boldsymbol{f}_{\boldsymbol{u}}{}^{\mathrm{T}}(t) \boldsymbol{\Phi}_o{}^{\mathrm{T}}(t_f,t) \left( w_H (H + \varphi_t) \left\{ H_{\boldsymbol{x}} + \varphi_{\boldsymbol{x}t} + \varphi_{\boldsymbol{xx}} \boldsymbol{f} \right\} \right) \Big|_{t_f} \end{aligned} \tag{63}$$

The EDE (57) on $t_f$ is modified with $\boldsymbol{n}_{t_f}$ being

$$\boldsymbol{n}_{t_f} = \left( \frac{1}{2} H_{\boldsymbol{u}}{}^{\mathrm{T}} H_{\boldsymbol{u}} + \left( \int_{t_0}^{t_f} \boldsymbol{\Phi}_o(t_f,s) \boldsymbol{f}_{\boldsymbol{u}} H_{\boldsymbol{u}} \mathrm{d}s \right)^{\mathrm{T}} \bar{L}_{\boldsymbol{x}} + w_H (H + \varphi_t)(\boldsymbol{f}^{\mathrm{T}} \varphi_{\boldsymbol{x}t} + \boldsymbol{f}^{\mathrm{T}} \bar{L}_{\boldsymbol{x}} + H_t + \varphi_{tt}) \right) \Bigg|_{t_f} \tag{64}$$

With the compact form evolution equation, the anticipated variable evolving along the variation time $\tau$, as depicted in Fig. 1, merely includes the control variables. Use the semi-discrete method to solve Eq. (56), only control variables are discretized for the numerical computation. Although both the compact form and the primary form evolution equations are derived from similar unconstrained functional, their concrete expressions are quite different. Besides the reduced dimensionality, it is shown that the EPDE (56) requires the integration, and the differentiation, as displayed in the EPDE (22), may be avoided. In seeking solutions, this is advantageous to reduce the numerical error.

## IV. Illustrative Examples

First a linear example taken from Xie [27] is solved.

**Example 1**: Consider the following dynamic system

$$\dot{\boldsymbol{x}} = \boldsymbol{A}\boldsymbol{x} + \boldsymbol{b}u$$

where $\boldsymbol{x} = \begin{bmatrix} x_1 \\ x_2 \end{bmatrix}$, $\boldsymbol{A} = \begin{bmatrix} 0 & 1 \\ 0 & 0 \end{bmatrix}$, and $\boldsymbol{b} = \begin{bmatrix} 0 \\ 1 \end{bmatrix}$. Find the solution that minimizes the performance index

$$J = \frac{1}{2}\int_{t_0}^{t_f} u^2 \mathrm{d}t$$

with the boundary conditions

$$\boldsymbol{x}(t_0) = \begin{bmatrix} 1 \\ 1 \end{bmatrix},\ \boldsymbol{x}(t_f) = \begin{bmatrix} 0 \\ 0 \end{bmatrix}$$

where the initial time $t_0 = 0$ and the terminal time $t_f = 2$ are fixed.

This example was solved with the primary form of the first evolution equation in Ref. [12]. Here we solve it with the compact form evolution equation, which was derived as

$$\frac{\partial u(t,\tau)}{\partial \tau} = -K\left\{u + \boldsymbol{b}^{\mathrm{T}}\boldsymbol{\lambda} + \boldsymbol{b}^{\mathrm{T}}\left(e^{\boldsymbol{A}(t_f - t)}\right)^{\mathrm{T}} \boldsymbol{W}_{\boldsymbol{x}_f}\boldsymbol{x}(t_f)\right\}$$

$$\frac{\mathrm{d}\boldsymbol{\pi}}{\mathrm{d}\tau} = -\boldsymbol{K}_{\boldsymbol{\pi}}\int_{t_0}^{t_f} e^{\boldsymbol{A}(t_f - t)}\boldsymbol{b}(u + \boldsymbol{b}^{\mathrm{T}}\boldsymbol{\lambda})\mathrm{d}t$$

The state and costate variables were computed by

$$\boldsymbol{x}(t,\tau) = e^{\boldsymbol{A}(t - t_0)}\begin{bmatrix} 1 \\ 1 \end{bmatrix} + \int_{t_0}^{t} e^{\boldsymbol{A}(t-s)}\boldsymbol{b}u(s,\tau)\,\mathrm{d}\,s$$

$$\boldsymbol{\lambda}(t,\tau) = \left(e^{\boldsymbol{A}(t_f - t)}\right)^{\mathrm{T}}\boldsymbol{\pi}(\tau)$$

The one-dimensional gain matrix $K$ was set as $K = 1$ and the $2\times 2$ dimensional gain matrix $\boldsymbol{K}_{\boldsymbol{\pi}}$ was $\begin{bmatrix} 1 & 0 \\ 0 & 1 \end{bmatrix}$. The weights were $\boldsymbol{W}_{\boldsymbol{x}_f} = \begin{bmatrix} 1 & \\ & 1 \end{bmatrix}$ and $w_H = 1$. The initial conditions were $u(t,\tau)\big|_{\tau=0} = 0$ and $\boldsymbol{\pi}\big|_{\tau=0} = \begin{bmatrix} 0 \\ 0 \end{bmatrix}$. Using the semi-discrete method, the time horizon $[t_0, t_f]$ was discretized uniformly with 41 points. Thus, a dynamic system with 43 states was obtained and the OCP was transformed to a finite-dimensional IVP. Note that in this discretization granularity, there will be 207 states for the primary form evolution equation. Huge difference. To solve the IVP, the ODE integrator "ode15s", with default relative error tolerance $1\times 10^{-3}$ and default absolute error tolerance $1\times10^{-6}$, was employed. The spline interpolation was used to obtain the control within the discretization points. For comparison, the analytic solution by solving the BVP is also presented.

$$
\begin{aligned}
\hat{x}_1 &= 0.5t^3 - 1.75t^2 + t + 1 \\
\hat{x}_2 &= 1.5t^2 - 3.5t + 1 \\
\hat{\lambda}_1 &= 3 \\
\hat{\lambda}_2 &= -3t + 3.5 \\
\hat{u} &= 3t - 3.5 \\
\hat{\boldsymbol{\pi}} &= \begin{bmatrix} 3 & -2.5 \end{bmatrix}^{\mathrm{T}}
\end{aligned}
$$

Fig. 2 shows the evolving process of $u(t)$ solutions towards the analytic. At $\tau$ =300s, the numerical solution is indistinguishable from the analytic, and this shows the effectiveness of the compact form evolution equation. In Fig. 3, the evolution profiles of the Lagrange multipliers are presented, and it is computed that $\boldsymbol{\pi}\big|_{\tau=300} = \begin{bmatrix} 3.0000 \\ -2.5000 \end{bmatrix}$, which equals the analytic result exactly. Figs. 4 and 5 plots the numerical solutions of the states and the costates at $\tau$ = 300s, respectively. It is shown that they are almost identical with the analytic solutions.

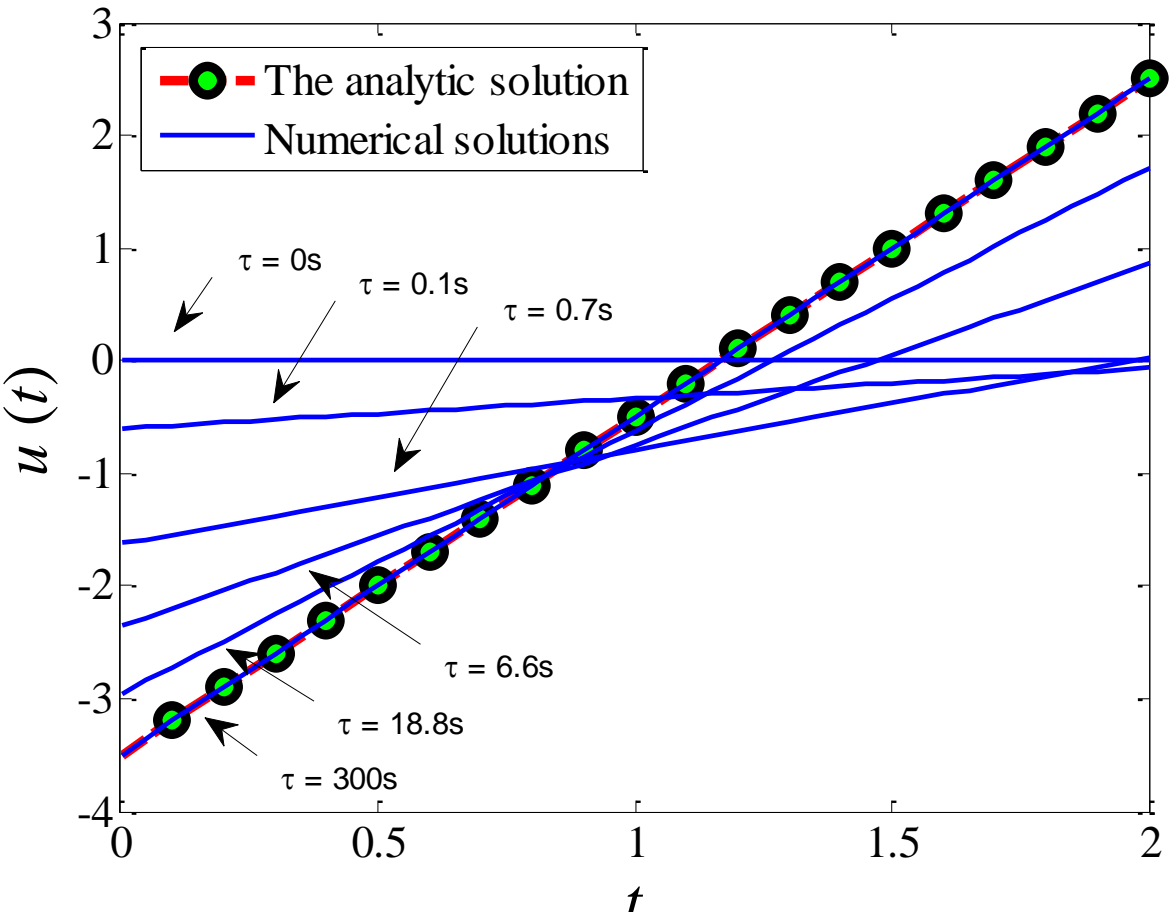

Fig. 2 The evolution of numerical solutions of $u$ to the analytic solution.

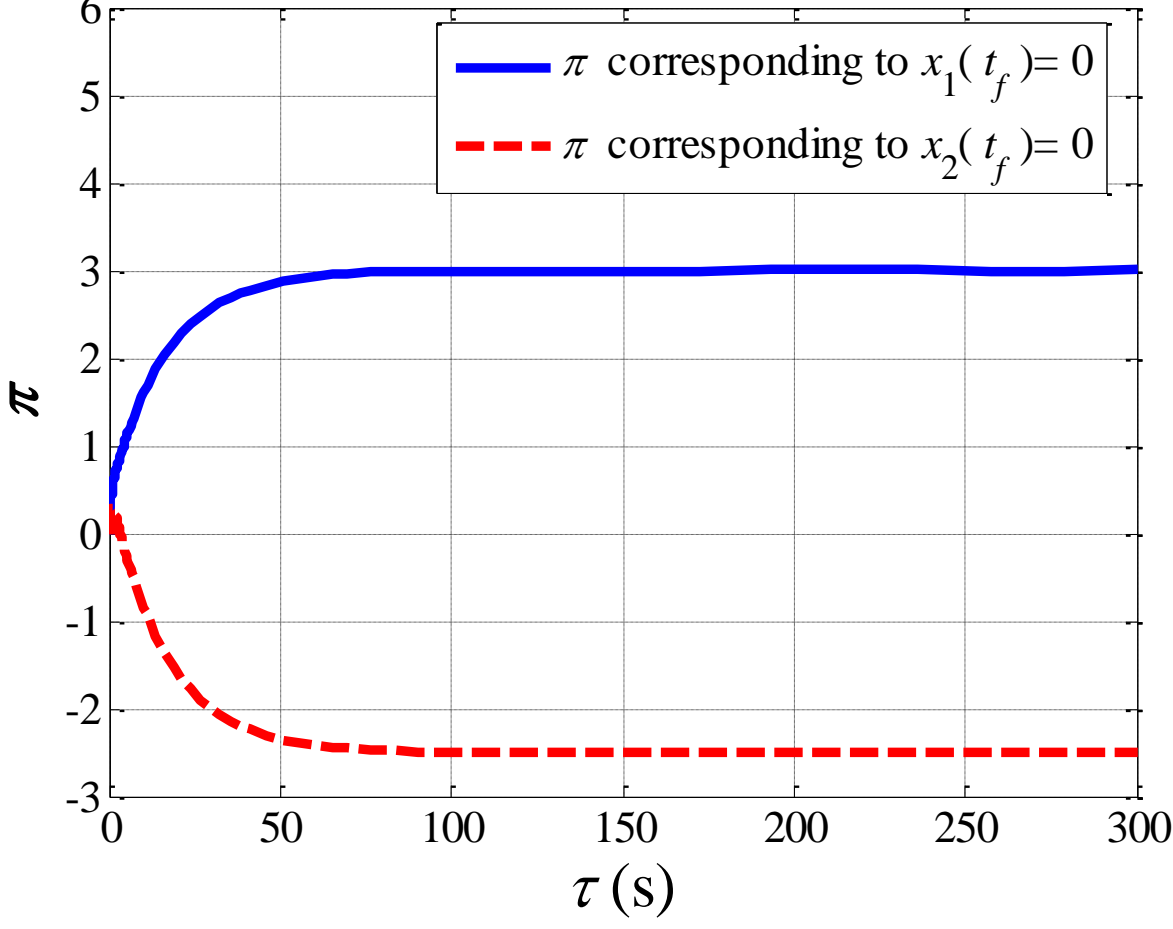

Fig. 3 The evolution profiles of Lagrange multipliers.

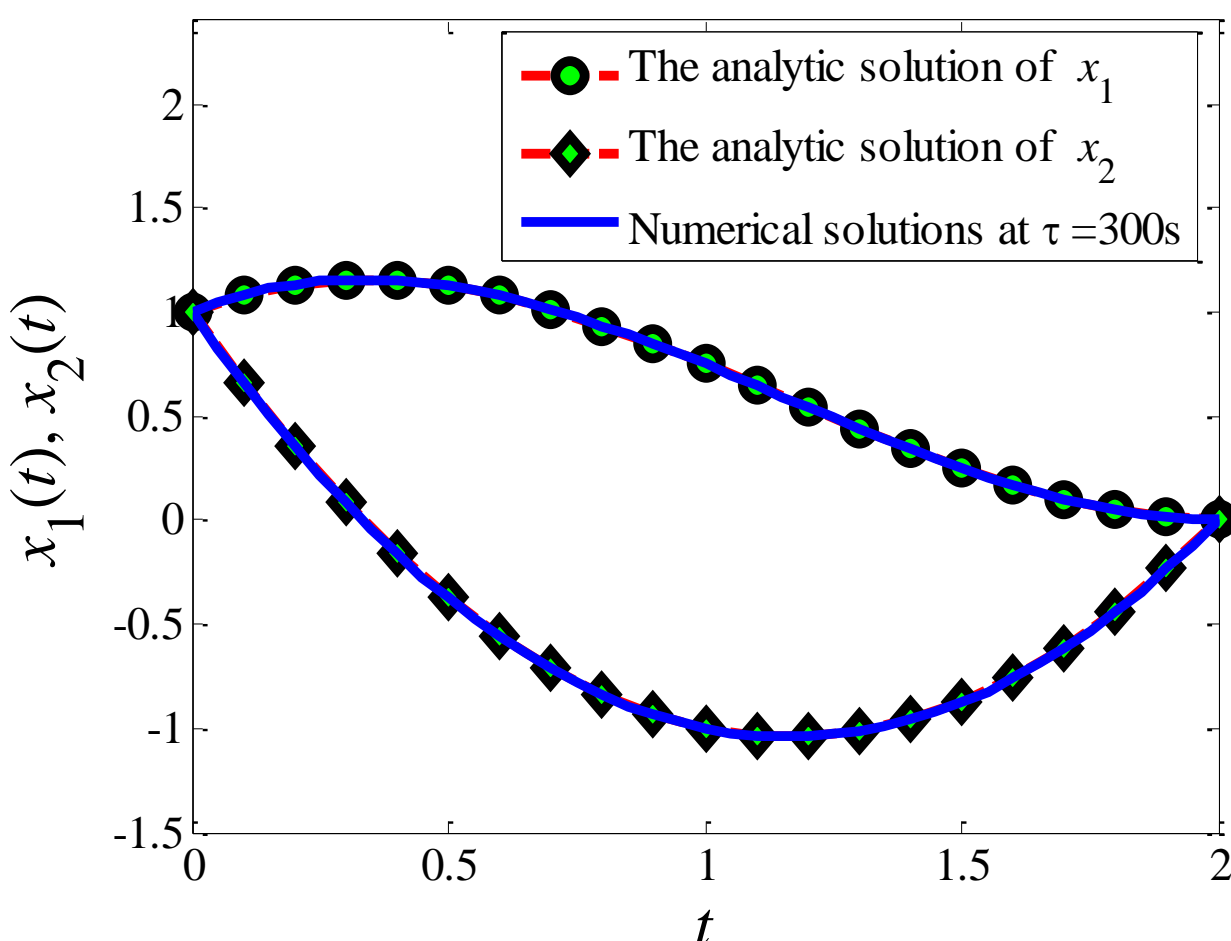

Fig. 4 The numerical solutions of $x_1$ and $x_2$ at $\tau = 300\text{s}$.

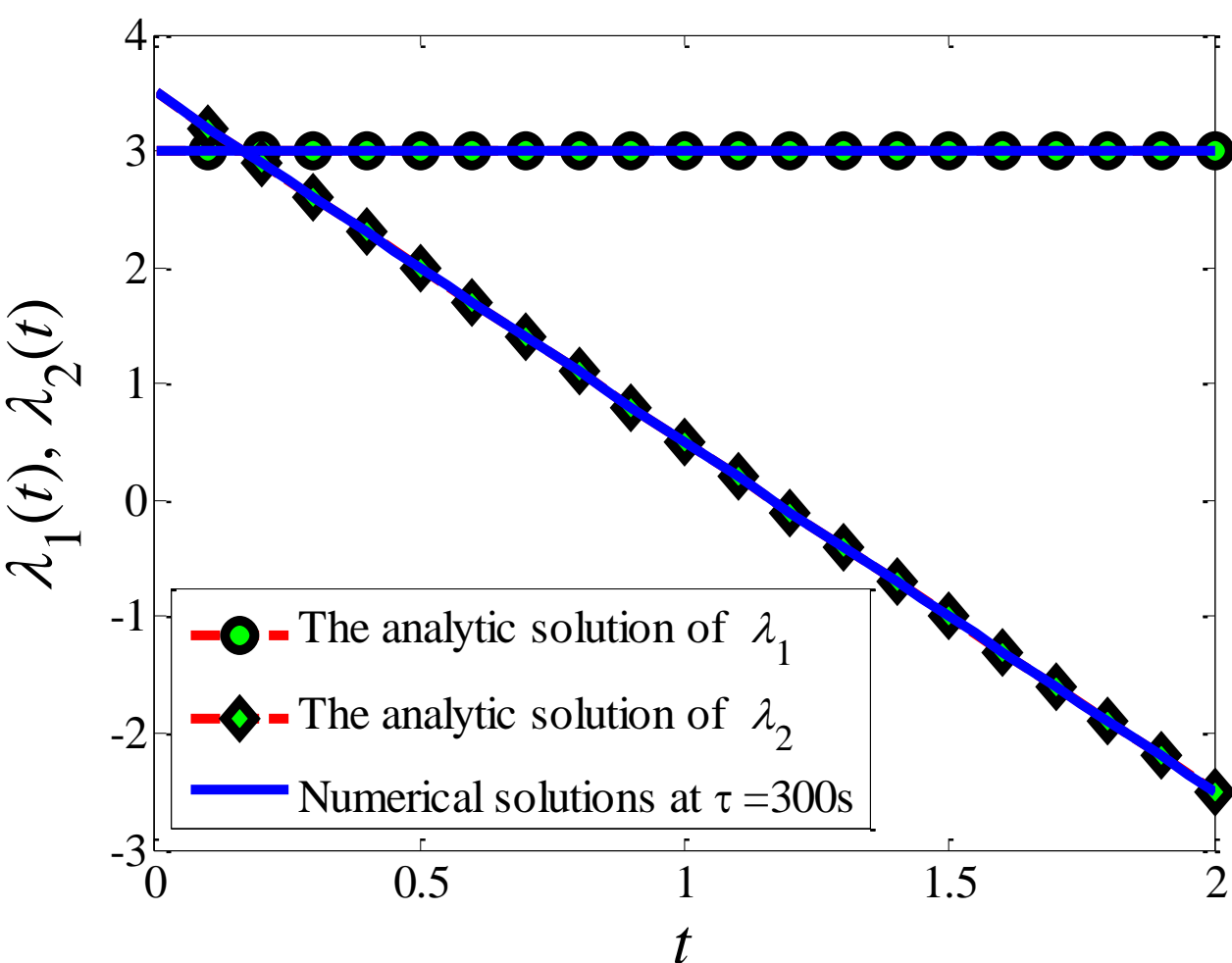

Fig. 5 The numerical solutions of $\lambda_1$ and $\lambda_2$ at $\tau = 300\text{s}$.

In Table 1, we compare the results with the compact form and primary form of the first evolution equations. The errors are denoted by $e_J = \left| J|_{\tau=300} - \hat{J} \right|$, $e_u = \max\limits_{t\in[t_0,t_f]} \left( \left| u(t,\tau)|_{\tau=300} - \hat{u} \right| \right)$, $e_{x_1} = \max\limits_{t\in[t_0,t_f]} \left( \left| x_1(t)|_{\tau=300} - \hat{x}_1 \right| \right)$, $e_{x_2} = \max\limits_{t\in[t_0,t_f]} \left( \left| x_2(t)|_{\tau=300} - \hat{x}_2 \right| \right)$, $e_{\lambda_1} = \max\limits_{t\in[t_0,t_f]} \left( \left| \lambda_1(t)|_{\tau=300} - \hat{\lambda}_1 \right| \right)$, and $e_{\lambda_2} = \max\limits_{t\in[t_0,t_f]} \left( \left| \lambda_2(t)|_{\tau=300} - \hat{\lambda}_2 \right| \right)$. It is shown that the precision is much higher with the compact form evolution equation, while it takes more time for the integration. Especially, in Ref. [20], we solved this example with the third evolution equation, a compact form as well. By comparison, it is found that between the two compact forms, the computation precision is similar while there is some difference in the computation time, on which the third evolution equation used shorter. This is mainly because of the less integration it required.

Table 1 Comparison of solutions at $\tau = 300\text{s}$ between the compact form and the primary form equations

| | Number of discretization points | Computation time | $e_J$ | $e_u$ | $e_{x_1}$ | $e_{x_2}$ | $e_{\lambda_1}$ | $e_{\lambda_2}$ |
|---|---|---|---|---|---|---|---|---|
| The compact form equation | 43 | 5.72 (s) | $6.86\times10^{-6}$ | $5.77\times10^{-6}$ | $7.65\times10^{-7}$ | $2.87\times10^{-6}$ | $5.74\times10^{-6}$ | $7.28\times10^{-6}$ |
| The primary form equation | 207 | 0.54 (s) | $2.45\times10^{-2}$ | $1.73\times10^{-2}$ | $4.32\times10^{-3}$ | $5.59\times10^{-3}$ | $1.61\times10^{-2}$ | $1.83\times10^{-2}$ |

Now we consider a nonlinear example with free terminal time $t_f$, the Brachistochrone problem [28], which describes the motion curve of the fastest descending.

**Example 2**: Consider the following dynamic system

$$\dot{\boldsymbol{x}} = \boldsymbol{f}(\boldsymbol{x}, u)$$

where $\boldsymbol{x} = \begin{bmatrix} x \\ y \\ V \end{bmatrix}$, $\boldsymbol{f} = \begin{bmatrix} V\sin(u) \\ -V\cos(u) \\ g\cos(u) \end{bmatrix}$, and $g = 10$ is the gravity constant. Find the solution that minimizes the performance index

$$J = t_f$$

with the boundary conditions

$$\begin{bmatrix} x \\ y \\ V \end{bmatrix}\Bigg|_{t_0=0} = \begin{bmatrix} 0 \\ 0 \\ 0 \end{bmatrix}, \begin{bmatrix} x \\ y \end{bmatrix}\Bigg|_{t_f} = \begin{bmatrix} 2 \\ -2 \end{bmatrix}$$

We solved this example with the primary form evolution equation in Ref. [12] before. Here it will be addressed via the compact form evolution equation. In the specific evolution equations (56)-(58), the gain parameters were set as $K = 0.1$, $k_{t_f} = 0.01$, and $\boldsymbol{K}_{\boldsymbol{\pi}} = \begin{bmatrix} 0.1 & \\ & 0.1 \end{bmatrix}$, respectively. The weights were $\boldsymbol{W}_{\boldsymbol{x}_f} = \begin{bmatrix} 1 & & \\ & 1 & \\ & & 1 \end{bmatrix}$ and $w_H = 1$. The initial conditions were set to be $u(t,\tau)\big|_{\tau=0} = 0$, $t_f(\tau)\big|_{\tau=0}$=1s and $\boldsymbol{\pi}\big|_{\tau=0} = \begin{bmatrix} 0 \\ 0 \end{bmatrix}$. We also discretized the time horizon $[t_0, t_f]$ uniformly, with 101 points. Then an IVP with 104 states is obtained. We employed "ode15s" to carry out the numerical integration for this transformed IVP with respect to $\tau$ while used "ode45" to solve the state equation and the costate equation with respect to $t$ in Eqs. (35) and (36). In the integrator setting, the relative error tolerance and the absolute error tolerance were $1\times10^{-3}$ and $1\times10^{-6}$, respectively. The spline interpolation was again used to get the control within the discretization points. For comparison, we computed the optimal solution with GPOPS-II [29], a Radau PS method based OCP solver.

The control solutions are plotted in Fig. 6, and the asymptotical approach of the numerical results are demonstrated. Fig. 7 presents the evolution profiles of the Lagrange multipliers $\boldsymbol{\pi}$. They reach the optimal value of $\begin{bmatrix} -0.1477 \\ 0.0564 \end{bmatrix}$ rapidly. Fig. 8 gives the states curve in the $xy$ coordinate plane, showing that the numerical results starting from the vertical line approach the optimal solution over time. In Fig. 9, the terminal time profile against the variation time $\tau$ is plotted. The result of $t_f$ declines rapidly at first and then approaches the minimum decline time gradually. It is almost unchanged after $\tau = 35$s. At $\tau = 400$s, we compute that $t_f = 0.8165$s, same to the result from GPOPS-II. In Table 2, the optimization results using the compact and the primary evolution equations respectively are also compared. Note that the optimal solutions from GPOPS-II are denoted by a hat " ^ ", $e_J = \left| t_f\big|_{\tau=400} - \hat{t}_f \right|$, $e_x = \max\limits_{t\in[t_0,t_f]} \left( \left| x(t)\big|_{\tau=400} - \hat{x} \right| \right)$, $e_y = \max\limits_{t\in[t_0,t_f]} \left( \left| y(t)\big|_{\tau=400} - \hat{y} \right| \right)$, $e_V = \max\limits_{t\in[t_0,t_f]} \left( \left| V(t)\big|_{\tau=400} - \hat{V} \right| \right)$, $e_{\lambda_x} = \max\limits_{t\in[t_0,t_f]} \left( \left| \lambda_x(t)\big|_{\tau=400} - \hat{\lambda}_x \right| \right)$, $e_{\lambda_y} = \max\limits_{t\in[t_0,t_f]} \left( \left| \lambda_y(t)\big|_{\tau=400} - \hat{\lambda}_y \right| \right)$, and $e_{\lambda_V} = \max\limits_{t\in[t_0,t_f]} \left( \left| \lambda_V(t)\big|_{\tau=400} - \hat{\lambda}_V \right| \right)$. It is again shown that the precision of solutions via the compact form evolution equation is much higher, and in this example the computation time it used is much smaller. By the way, recall the results from solving the third evolution equation in Ref. [20], the difference is also mainly on the computation time while the precision is similar.

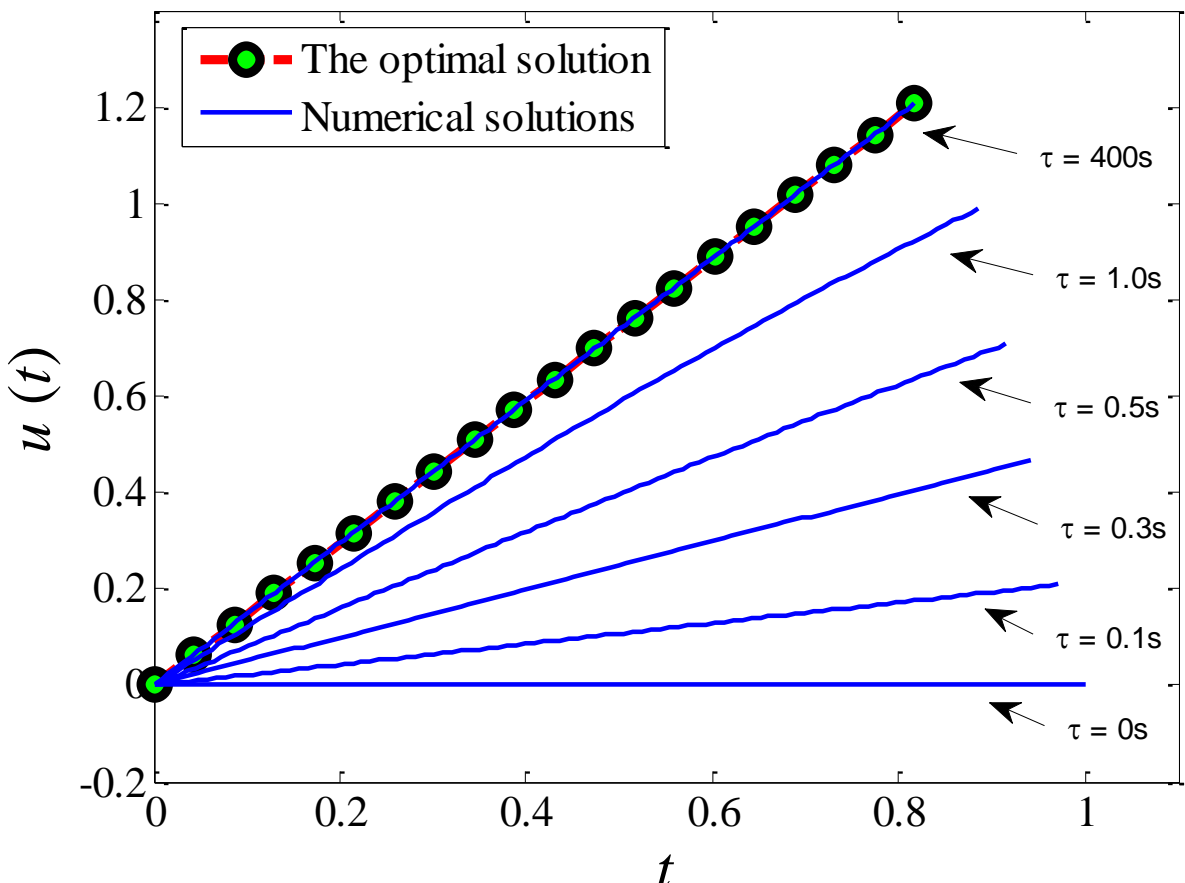

Fig. 6. The evolution of numerical solutions of $u$ to the optimal solution.

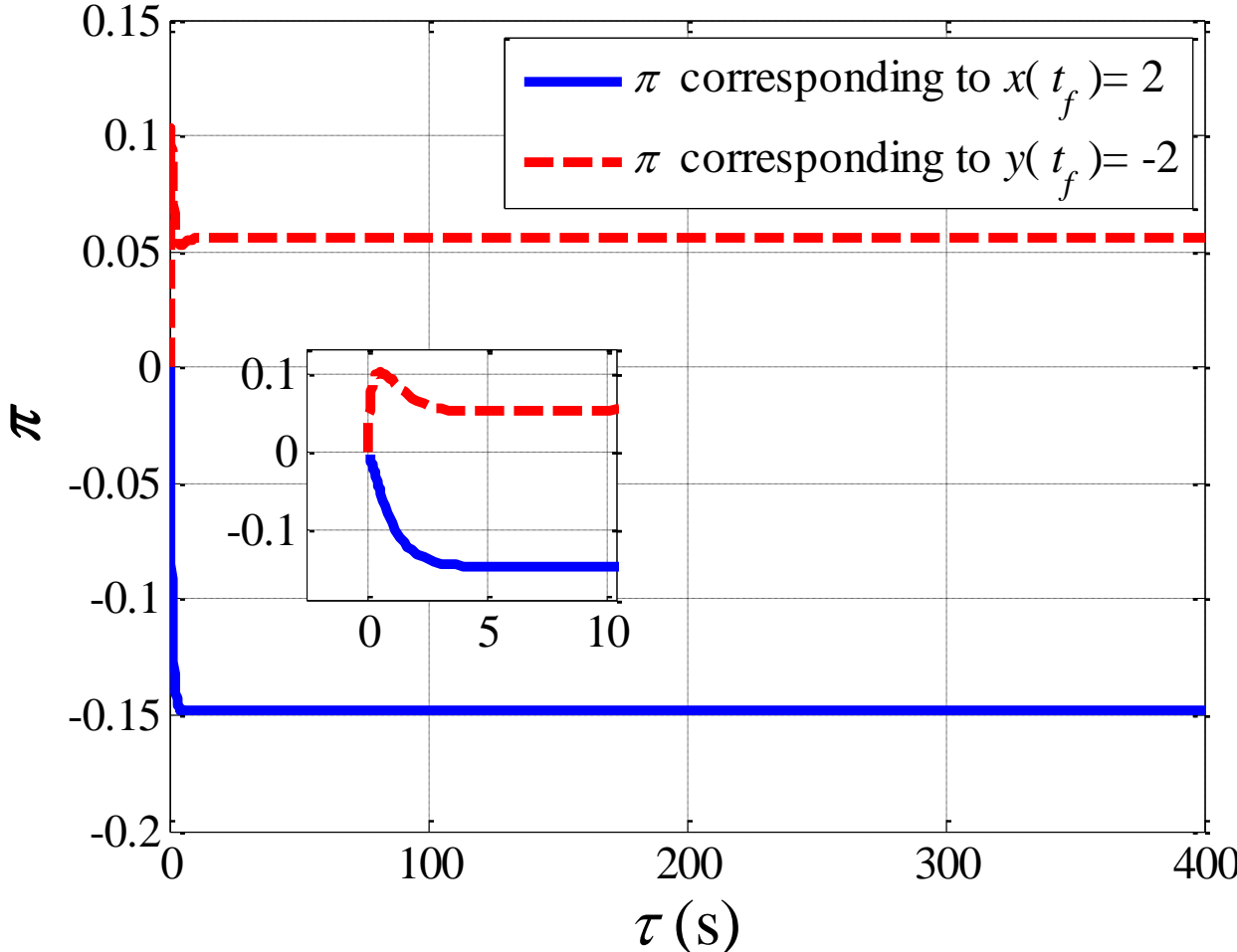

Fig. 7 The evolution profiles of Lagrange multipliers.

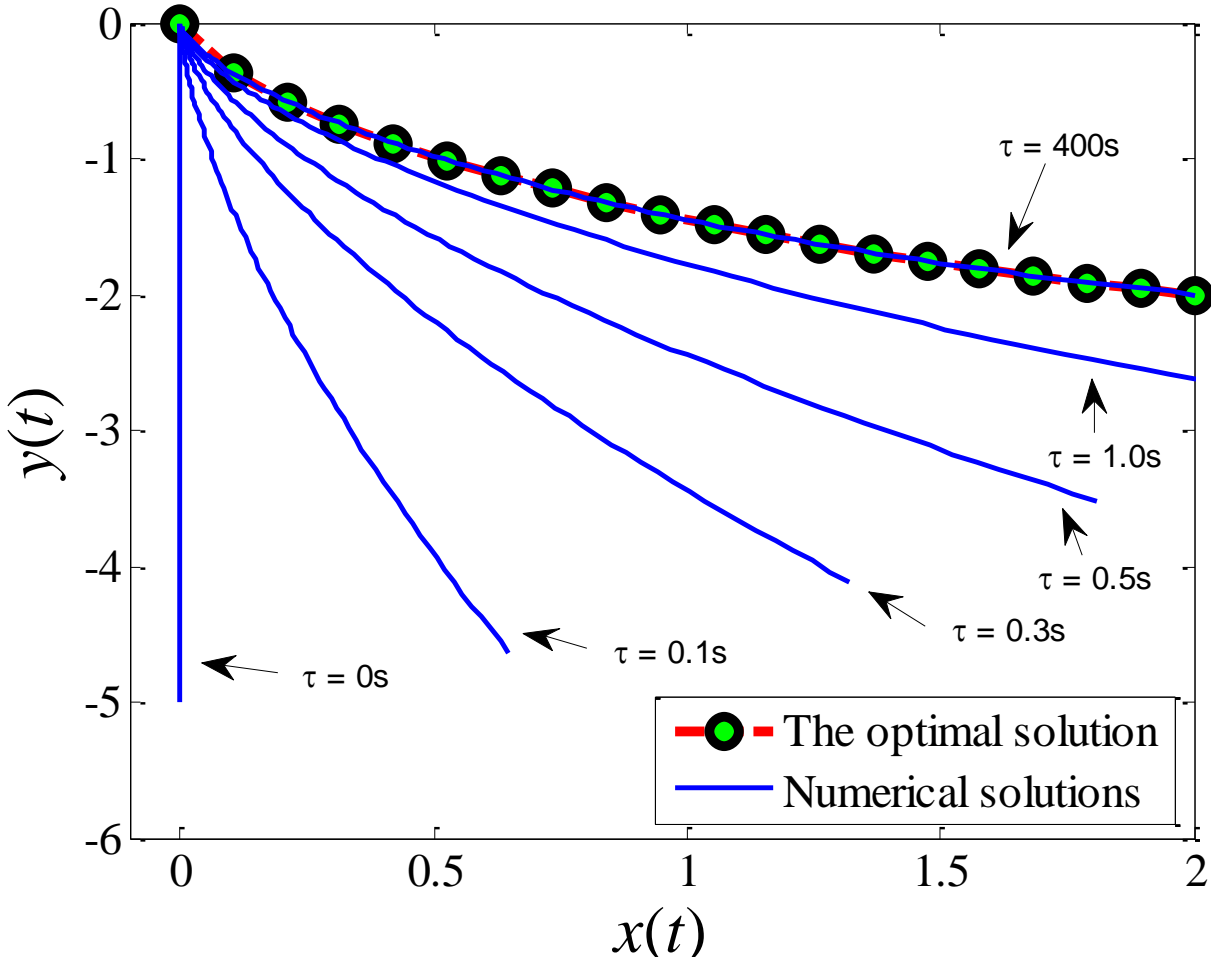

Fig. 8 The evolution of numerical solutions in the $xy$ coordinate plane to the optimal solution.

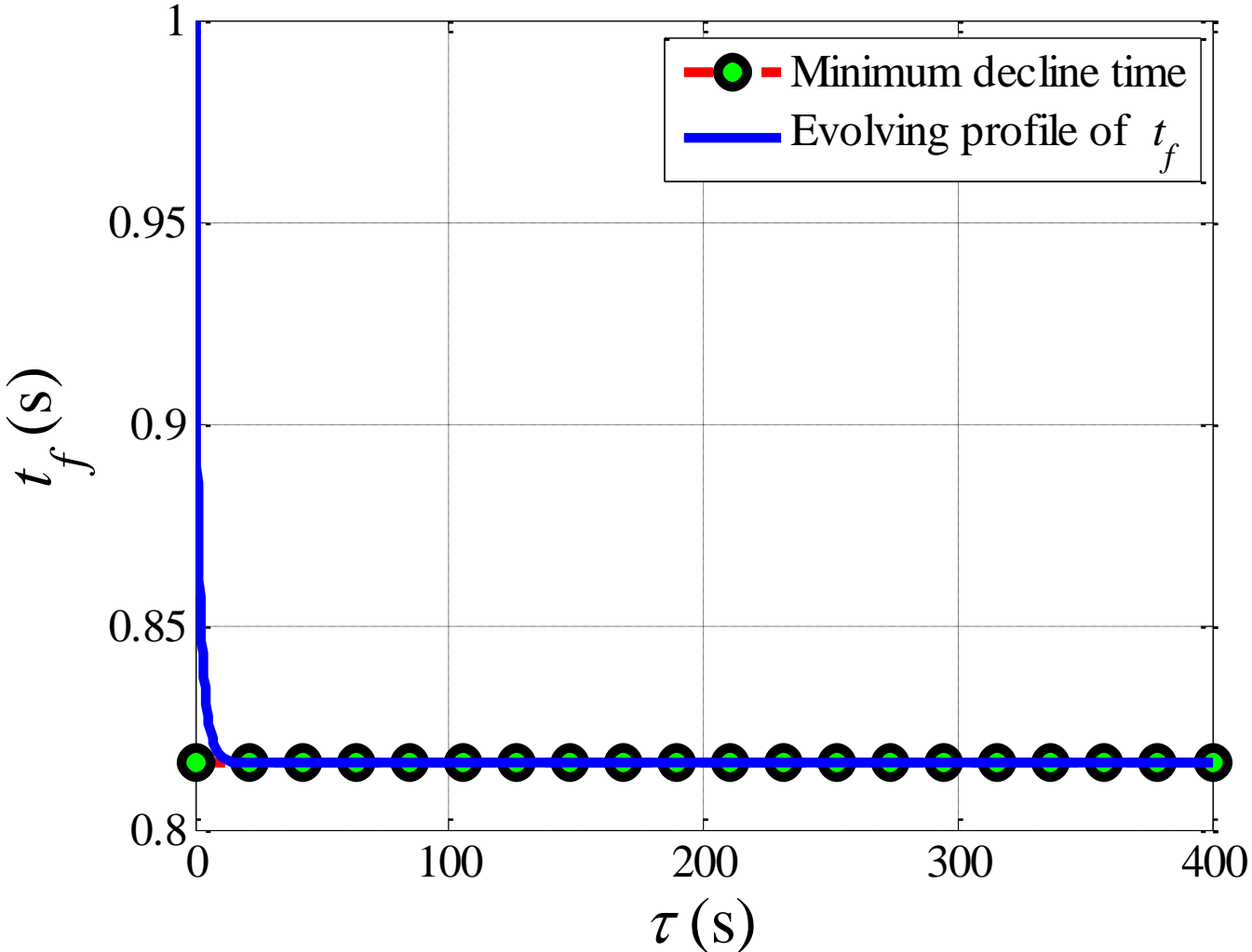

Fig. 9 The evolution profile of $t_f$ to the minimum decline time.

Table 2 Comparison of solutions at $\tau = 400\text{s}$ between the compact form and the primary form equations

| | Number of discretization points | Computation time | $e_J$ | $e_u$ | $e_x$ | $e_y$ | $e_V$ | $e_{\lambda_x}$ | $e_{\lambda_y}$ | $e_{\lambda_V}$ |
|---|---|---|---|---|---|---|---|---|---|---|
| The compact form equation | 104 | 56.24 (s) | $7.51\times 10^{-10}$ | $2.74\times 10^{-4}$ | $1.79\times 10^{-5}$ | $2.25\times 10^{-5}$ | $3.79\times 10^{-5}$ | $7.60\times 10^{-7}$ | $7.60\times 10^{-7}$ | $9.24\times 10^{-7}$ |
| The primary form equation | 710 | 239.62 (s) | $9.82\times 10^{-5}$ | $4.60\times 10^{-3}$ | $1.14\times 10^{-3}$ | $2.16\times 10^{-3}$ | $4.10\times 10^{-3}$ | $1.01\times 10^{-4}$ | $2.53\times 10^{-4}$ | $2.92\times 10^{-4}$ |

## V. CONNECTIONS TO THE CLASSIC ITERATION METHODS

In this section, the relations between the VEM and the classic numerical iteration methods that solve OCPs will be discussed. We will analogize them by considering a parameter optimization problem with performance index

$$J = h(\boldsymbol{\theta}) \tag{65}$$

where $\boldsymbol{\theta} \in \mathbb{R}^s$ is the optimization parameter vector and $h:\mathbb{R}^s \to \mathbb{R}$ is a scalar function. To find the optimal value $\hat{\boldsymbol{\theta}}$ that minimizes $J$, the classic gradient method uses the iteration scheme that reduces $J$ with update of $\boldsymbol{\theta}$, that is

$$\boldsymbol{\theta}_{k+1} = \boldsymbol{\theta}_k - \alpha\, h_{\boldsymbol{\theta}}\big|_{\boldsymbol{\theta}_k} \tag{66}$$

where $\alpha$ is a positive step parameter, $h_{\boldsymbol{\theta}} = \dfrac{\partial h}{\partial \boldsymbol{\theta}}$ is the gradient and the subscript $k$ denotes the number of iteration. In another way, the Newton method, which is developed upon the second-order approximation, solves the problem with

$$\boldsymbol{\theta}_{k+1} = \boldsymbol{\theta}_k - \beta h_{\boldsymbol{\theta}\boldsymbol{\theta}}{}^{-1} h_{\boldsymbol{\theta}}\big|_{\boldsymbol{\theta}_k} \tag{67}$$

where $\beta$ is a positive step parameter and $h_{\boldsymbol{\theta}\boldsymbol{\theta}} = \dfrac{\partial^2 h}{\partial \boldsymbol{\theta}^2}$ is the Hessian matrix. For both Eqs. (66) and (67), the iteration will halt when $\boldsymbol{\theta}$ reaches $\hat{\boldsymbol{\theta}}$, which satisfies $h_{\boldsymbol{\theta}}\big|_{\boldsymbol{\theta}=\hat{\boldsymbol{\theta}}} = \boldsymbol{0}$, i.e., the first-order optimality condition. A general expression for these numerical iteration schemes may be

$$\boldsymbol{\theta}_{k+1} = \boldsymbol{M}_{\boldsymbol{\theta}}\left[\boldsymbol{\theta}_k\right] \tag{68}$$

where $\boldsymbol{M}_{\boldsymbol{\theta}}$ represents the mapping operator associated with some optimization method. After each iteration, we get a new estimate of the solution based on the discrete dynamics.

Now consider the continuous-time dynamics way to solve this problem. We regard $J$ , i.e., function $h$ here, as the Lyapunov function. By Differentiating it with respect to a virtual time $\tau$ , which is used to describe the derived dynamics, we have

$$\frac{\mathrm{d}J}{\mathrm{d}\tau} = {h_{\boldsymbol{\theta}}}^{\mathrm{T}} \frac{\mathrm{d}\boldsymbol{\theta}}{\mathrm{d}\tau} \tag{69}$$

To guarantee that $J$ decreases with respect to $\tau$ , i.e., $\frac{\mathrm{d}J}{\mathrm{d}\tau} \leq 0$ , we may set

$$\frac{\mathrm{d}\boldsymbol{\theta}}{\mathrm{d}\tau} = -k_g h_{\boldsymbol{\theta}} \tag{70}$$

where $k_g$ is a positive scalar. Under such dynamics, $h$ will decrease until it reaches a minimum, and $\boldsymbol{\theta}$ will approaches $\hat{\boldsymbol{\theta}}$ asymptotically to meet the first-order optimality condition. Consider another way. Use the fist-order optimality condition to construct a Lyapunov function candidate as

$$\bar{J} = {h_{\boldsymbol{\theta}}}^{\mathrm{T}} h_{\boldsymbol{\theta}} \tag{71}$$

Similarly we would have

$$\frac{\mathrm{d}\bar{J}}{\mathrm{d}\tau} = 2{h_{\boldsymbol{\theta}}}^{\mathrm{T}} h_{\boldsymbol{\theta\theta}} \frac{\mathrm{d}\boldsymbol{\theta}}{\mathrm{d}\tau} \tag{72}$$

To ensure that $\bar{J}$ decreases, we may establish that

$$\frac{\mathrm{d}\boldsymbol{\theta}}{\mathrm{d}\tau} = -k_n {h_{\boldsymbol{\theta\theta}}}^{-1} h_{\boldsymbol{\theta}} \tag{73}$$

or develop the inversion-free version as

$$\frac{\mathrm{d}\boldsymbol{\theta}}{\mathrm{d}\tau} = -k_n {h_{\boldsymbol{\theta\theta}}}^{\mathrm{T}} h_{\boldsymbol{\theta}} \tag{74}$$

where $k_n$ is a positive scalar. Expectedly, Eqs. (73) and (74) will seek the solution that satisfies $h_{\boldsymbol{\theta}} = \boldsymbol{0}$ . By comparison, we may find that Eqs. (66) and (67) are actually the discrete implementation of Eqs. (70) and (73).

In solving the OCP, the mechanism that the classic iteration methods use may be generally described as

$$\left(\boldsymbol{u}(t)\right)_{k+1} = \boldsymbol{M}_{\boldsymbol{u}}\left[\left(\boldsymbol{u}(t)\right)_{k}\right] \tag{75}$$

where $\boldsymbol{M}_{\boldsymbol{u}}$ represents the mapping operator associated with some optimization method. A vital concept introduced in the VEM is the variation time $\tau$ . At first glance, it seems strange and might not be easily accepted. However, if interpreted from the view of numerical computation dimension, it actually exists long. Namely, the numerical iterations are just the reflection of discrete dynamics along the virtual time dimension $\tau$ . Recall the gradient method that solves the OCP [26], its iteration mechanism is actually the discrete version of the third evolution equation developed in Ref. [20], while the compact form of the first evolution equation proposed in this paper may be regarded as a continuous realization of the Newton type iteration mechanism. Intuitively, any discrete iteration method may have its continuous-time evolving counterpart. In the continuous form, daunting task of searching reasonable step size and annoying oscillation phenomenon around the optimum are eliminated and the mature Ordinary Differential Equation (ODE) integration methods may be employed to solve the OCPs conveniently.

## VI. Conclusion

The compact form of the first evolution equation is formulated for better performance. It only considers the evolution of the control variables along the virtual variation time, thus decreasing the dimension of the Evolution Partial Differential Equation (EPDE) in the primary form. Even with arbitrary initial conditions, its solution is guaranteed to ultimately meet the optimality conditions by the Lyapunov principle. Notably, the compact form evolution equation employs the integral instead of the differential in the right function, and this is advantageous in seeking solutions. Using the semi-discrete method to solve this equation, the scale of the transformed Initial-value Problem (IVP) is significantly reduced. From the illustrative examples, it is shown that the compact form outperforms the primary form in both the precision and the efficiency for the dense discretization. In particular, by comparison with the classic numerical iteration methods, we establish the connections that the computation scheme in the gradient method is the discrete implementation of the third evolution equation proposed previously, and the compact form of the first evolution equation is a continuous version of the Newton type iteration mechanism.